\documentclass[letterpaper, a4paper]{amsart}

\usepackage{amsmath}
\usepackage{latexsym}
\usepackage[all]{xy}
\usepackage{color}
\newtheorem{teorema}{Theorem}[section]

\include{amslatex}

\newtheorem{comentario}[teorema]{Remark}

\numberwithin{equation}{section}

\begin{document}
\begin{title}[Quantum Radar]
 {Introduction to quantum radar}
\end{title}
\date{\today}
\maketitle
\thispagestyle{empty}
\begin{center}
\author{Ricardo Gallego Torrom\'e\footnote{Email: rigato39@gmail.com}}
\end{center}
\begin{center}
\address{Department of Mathematics\\
Faculty of Mathematics, Natural Sciences and Information Technologies\\
University of Primorska, Koper, Slovenia}
\end{center}
\bigskip
\begin{center}
\author{Nadya Ben Bekhti-Winkel\footnote{Email: nadya.ben.bekhti-winkel@fhr.fraunhofer.de}}
\end{center}
\begin{center}
\address{FHR, Fraunhofer-Institut f\"ur Hochfrequenzphysik und Radartechnik, Fraunhoferstr. 20, 53343 Wachtberg, Germany}
\end{center}
\bigskip
\begin{center}
\author{Peter Knott\footnote{Email: peter.knott@fhr.fraunhofer.de}}
\end{center}
\begin{center}
\address{FHR, Fraunhofer-Institut f\"ur Hochfrequenzphysik und Radartechnik, Fraunhoferstr. 20, 53343 Wachtberg, Germany}
\end{center}
\begin{center}
\address{Chair of Radar Systems Engineering
Institute of High Frequency Technology
RWTH Aachen University
Melatener Straße 25
52074 Aachen
Germany}
\end{center}
\begin{abstract}
After a brief introduction to the notion of quantum entanglement and quantum correlations, several schemes for a quantum radar based upon the quantum illumination and others protocols are discussed. We review different concepts that have been introduced to overcome several of the inherent difficulties in the implementation of quantum generation and/or detection quantum sensing protocols for RADAR applications. Our review is an up-to date critical presentation of the state of the art, with emphasis in the case by case assessment of the feasibility of the different concepts. We also aim that the review is accessible to non-experts in the field. Hence several appendixes and a technical glossary are included.
\end{abstract}
\newpage
\tableofcontents
\newpage
\section{Introduction}
After preliminary results of M. Sacchi on entanglement enhancement in entanglement
breaking quantum channels \cite{Sacchi2005a,Sacchi2005b} and the introduction of the quantum Chernoff bound \cite{Audenaert et al., Pirandola Lloyd}, S. Lloyd devised the protocol of quantum illumination \cite{Lloyd2008}
for discrete variable systems and Tan et al. \cite{Tan} cite the corresponding version for continuous variable systems. Quantum illumination is an example of quantum protocol where the enhancement in sensitivity derived by the use of quantum entangled states survives the effects of environmental noise. Indeed, the advantage of entanglement respect to non-entangled sources is, surprisingly and counter-intuitively, larger when the system is immersed in noisy environments.

This theoretical results ignited several investigations and demonstrations of quantum illumination, at the theoretical and experimental level. One of the goals in such research was and still remains, the realization of a {\it quantum radar}, a radar that explodes quantum entanglement, quantum illumination being a pre-eminent, to target range detection with sensitivity beyond classical radar schemes. As a result of this intensive research activity, it became clear that the enhancement of quantum illumination over classical illumination was limited by theoretical and technological considerations and that the goal of the realization of a quantum radar will need to wait the resolution of several issues at the practical and theoretical levels and also, that the enhancement cannot not be as strong as initially was thought. Nevertheless, even the more modest theoretical advantages that quantum illumination had over classical illumination for target detection remained a strong  motivation to develop the concept and more generally, to study quantum radar protocols and its practical and technological implementations. The major goal was the realization of a quantum radar, a goal which has not been reached yet. But other more modest potential applications has raised during these years.

The present review aims to introduce to the non-expert in quantum radar the main achievements, concepts, methods and results of the field, but also to explain the limitations and problems that appear in the theory and in its experimental and practical technological implementations. The pre-requisites for understanding the material presented here is to have a basic acquaintance with the fundamental mathematical tools of quantum mechanics (notion of Hilbert space and linear operators, scalar products,annihilation and creation operators,...). We expect that the review serves as a bridge between two communities. First, the quantum radar community, as a research community in the interface between  quantum optics, quantum metrology and quantum sensing. For them we tried to present an accurate, complete and realistic survey of the field. Second, the classical radar community with interest to initiate research in the area of quantum radar. For them we have tried to present a basic and clear exposition of the field.
We have approached the review with a minimal of formalism and we have developed several notions of quantum mechanics, quantum information and quantum optics in the appendix section and in the Glossary. In this way the reader familiar with the more technical notions can be dispensed from following their reading in the main part of the manuscript.

The structure of this review is the following. In section \ref{quantum entanglement and quantum radars}, a succinct introduction to quantum entanglement and several aspects of entanglement of relevance for quantum radar applications is provided. Then a general description of several approaches to quantum radar is presented. Section \ref{quantum interferometric radar protocols} describes quantum interferometric radar. Aspects of quantum illumination are discussed in sections \ref{Quantum Illumination} and \ref{problems of the microwave qi}. Section \ref{Hybrid protocols} discusses experimental implementations of quantum illumination where the receiver uses classical digital techniques for the process of the signal. Section \ref{Maccone Ren protocol} is an introduction to a novel approach to quantum radar developed by L. Maccone and C. Ren \cite{MacconeRen2019}. Section \ref{quantum illumination with multiple entangled photons} describes an innovative approach to quantum radar based in a combination of an idea extracted from  Maccone-Ren protocol  and Lloyd's quantum illumination. Section \ref{Other protocols} describes other new approaches to quantum radar, especially a recent proposal by Durak et al. that uses only classical concepts but illumination with entangled light \cite{DurakJamDindar2019}.
We observe that several of the developments in the appendix are not easy  to be found directly in the literature. This applies to exposition of several details of Lloyd's quantum illumination that we have presented in Appendix \ref{AppendixLloyd quantum illumination}. The Glossary introduces several technical terms that are of relevance for quantum illumination and quantum radar. The Appendix \ref{Quantum states for quantum radar} introduce most of the relevant states used in quantum radar. Appendix \ref{non-linear optics section} introduces several non-linear quantum optics processes of interested for quantum radar.

We are intellectually in debt with several reviews on quantum illumination and quantum radar that have appeared recently in the literature. Especially, the review on quantum illumination by J. Shapiro \cite{Shapiro2019} and the book on quantum radar by M. Lanzagorta \cite{Marco Lanzagorta 2011}. From the first, we depart mainly in the way of addressing the presentation and in the scope of the topics considered, since \cite{Shapiro2019} only discusses quantum illumination protocols. Other recent reviews that present a critical discussion on quantum illumination and that we have found very useful are \cite{Pirandola et al. 2018} and \cite{Sorelli et al.}.  From the second, we depart in an more up to date account (see also the special issue of IEEE magazine dedicated to quantum radar \cite{Fras ca Farina Balaji}). We are in debt with S. Pirandola for several very valuable remarks. We also would like to thank to R. di Candia, S. Paraoanu, L. Maccone, Changliang Ren and Q. Zhuang for comments and correspondence. Despite these influences, the authors take any resposability for possible mistakes or inaccuracies in this review paper.

 Finally, a review of this nature can hardly be a comprehensive one. Most of the topics are forced by the main research lines, but others are emphasized from the particular interests of the authors. Moreover, the field of quantum radar is in  continuous evolution and probably several relevant contributions have been not mention directly. We apologize if such research is not appropriately considered in this review. On the other hand, we hope that the review is useful for the researchers willing to enter this new research line, either from a theoretical point of view or more practical inclinations.

\section{Quantum entanglement, non-classical correlations and types of quantum radars}\label{quantum entanglement and quantum radars}
All quantum radar protocols are built and exploit non-classical properties of electromagnetic radiation, namely, quantum entanglement and quantum correlations. According with the characteristics and use of entanglement and related associated non-classical correlations,
quantum sensors are adequately denoted to be of {\it Type 1}, {\it Type 2} or {\it Type 3} \cite{Harris 2009, Marco Lanzagorta 2011}:
\begin{itemize}
\item {\bf Type 1 quantum sensors}: The quantum sensor transmits un-entangled, non-coherent quantum states of light. This type includes single photons quantum radars and classical LIDARS. An example of a Type-1 sensor is the LIDAR proposed in \cite{Feng Fei et al. 2017}, where squeezed light is used for detection.

\item {\bf Type 2 quantum sensors}: The quantum sensor transmits coherent light (classical states of light), but uses quantum photo sensors to increase performance in detection. References \cite{Harris 2009, Marco Lanzagorta 2011} offer different scope in the treatment of quantum enhanced LIDARS. An extensive treatment of this type of sensors can be found in \cite{Harris 2009}.

\item {\bf Type 3 quantum sensors}: The quantum sensors of this type transmit quantum states of light which are originally entangled, usually with the receiver, but the signal beam is not entangled with itself.

\end{itemize}
Recently, a new type of protocols have been discussed in the literature:
\begin{itemize}
\item {\bf Type 4 quantum sensors}: The signal beam is composed by light states that are entangled with itself. Maccone-Ren protocol \cite{MacconeRen2019} is of this type, as well the quantum illumination protocol that uses quantum states with multiple entangled photons \cite{Ricardo2020b}.
\end{itemize}

In this review we will concentrate our attention in  protocols for type 3 and type 4 quantum sensors. Type 1 and type 2 were extensively considered in \cite{Harris 2009}.

Quantum illumination and quantum radar refer to quantum sensing protocols based on the use of entangled sources of photon beams for radar applications and other quantum sensing and metrological applications. In such protocols, the use of entanglement varies from the application of a strong notion of entanglement, where the entanglement is preserved up to the reception of the scattered signal, to the use of entangled breaking discrimination protocols, where although entanglement is not present between the idler and the received signal, the systems still present certain enhancement in sensitivity  due to the inherited non-classical correlation properties of the initially entangled states. The disappearance of entanglement is caused by attenuation and environmental decoherence totally degrades the quantum coherence before the signal arrives to the detector.

Since these notions of quantum entanglement and the related notion of entanglement breaking channel are necessary to understand quantum radar protocols and quantum illumination protocols, a brief introduction to them and few related notions is in order. Although the concepts and applications are counter-intuitive at some point, in the present review, a pragmatic point of view has to be adopted. Hence the concepts that we will discuss related with quantum entanglement and quantum correlations pertain to the ambit of standard quantum mechanics, quantum optics and quantum information.
\subsection{Quantum entanglement}
 An  introduction to the concept of entanglement  is necessary to understand its applications in quantum illumination and quantum radar. An adequate introduction for this purpose can be found in chapter 6 of the monograph of Garrison and Chiao \cite{GarrisonChiao}. A comprehensive  treatment of quantum entanglement for continuous states  of entanglement can be found in the review paper  \cite{AdessoIlluminati2007}. Reference \cite{GarrisonChiao} also offers an introduction to the notions of quantum optics need for quantum illumination and quantum radar. Suitable reviews on quantum metrology are \cite{GiovannettiLloydMaccone, Pirandola et al. 2018}.

Quantum entanglement is a physical property consequence of the fundamental principle of linear superposition of states in quantum mechanics applied to compound systems.
The most elementary notion of quantum entanglement describes systems composed by two or more distinguishable parts , is introduced in the following lines. For simplicity, let us consider that the system is composed by two parts, here denoted by $1$ and $2$.
If the Hilbert space\footnote{Note that dynamical systems described by first order differential equations can be formulated using Hilbert space theory, by means of Koopman-von Neumann theory \cite{Koopman1931, Von Neumann}. This can lead to the notion of {\it classical entanglement}.} describing the system is a product of the form $\mathcal{H}=\,\mathcal{H}_1\otimes \,\mathcal{H}_2$, the separable  elements of $\mathcal{H}$ are the ones compatible with the {\it classical principle of separability}:

\bigskip
{\it Complete knowledge of the state of a compound systems yields complete knowledge of the individual states of the parts $1$ and $2$.}
\bigskip

The origin of this principle goes back to the discussion of the Einstein-Podolsky-Rosen paradox \cite{EinsteinPodolskiRosen} and was formulated by E. Schr\"odinger \cite{Schroedinger1935}. The negation of this postulate leads to the notion of entanglement. Therefore, an entangled state is such that the maximal knowledge of the compound states does not imply a complete knowledge of individual states of the parts. Since the maximal knowledge of a quantum (pure) state is provided by the wave function, the state is entangled if the wave function of the total system do no determine the wave functions of the compounds.

 The notion of quantum entanglement is without doubt, among the key notions in modern applications of quantum physics in  technology.
 The theory of quantum computation is based upon the quantum mechanical notion of entanglement, its formal properties and the possibilities of physical realization. All the quantum radar protocols that we will consider in this review are based on the concept of entanglement. On the other hand, quantum entanglement is probably (al-together with the notion of quantum non-locality) among the most difficult concepts to grasp in quantum theory from a spacetime perspective. Attempts to understand the concept in a geometric way have been recently explored. Two examples are the $EP=EPR$ conjecture \cite{MaldacenaSusskind, Susskind 2014} and two-dimensional time dynamics (see for instance \cite{Ricardo2017b} and chapter 9 in \cite{Ricardo2020}). Although advance on quantum foundations could bring changes in the description and understanding of quantum entanglement and quantum correlations, in this review we adopt a conventional point of view on quantum entanglement and quantum correlations, with the view on applications of the theory in quantum radar.

In the next paragraphs we will discuss several forms of entanglement and several associated notions of quantum correlations associated to entanglement which  are of relevance in different quantum radar protocols.

\subsection{Entanglement in pure states} Separable pure states in quantum mechanics are described by elements of the Hilbert space $\mathcal{H}$ that are compatible with the classical principle of separability. They are elements  $\psi\in\,\mathcal{H}_1\otimes \mathcal{H}_2$  that can be described as product of the form  $\psi=\,\psi_1\otimes \psi_2$, with $\psi_1\in\,\mathcal{H}_1$ and $\psi_2\in\,\mathcal{H}_2$.
For this class of separable states, the statistical properties associated to the subsystem $1$ and the statistical properties associated to the subsystem $2$ are statistically independent from each other. This can be seen as a consequence from the fact that for separable states, the joint density probability function (in the coordinate basis) is the product
\begin{align*}
|\psi(x_1,x_2)|^2=\,|\psi_1(x_1)|^2\,|\psi_2(x_2)|^2,
\end{align*}
 where $x_1$ labels the components of $\psi_1$ and $x_2$ the components of $\psi_2$. Remarkably, according to quantum mechanics, $\psi$, $\psi_1$ and $\psi_2$ are states from which we can extract the maximal quantum mechanical information, that is, all the  statistical information on measurements of observables when experiments are performed with the total system $1\sqcup 2$, the system $1$ and the system $2$, respectively. Also, the information for $1$ (respectively, respect to $2$) is obtained from the state $\psi$ by the process of integrating respect to $2$ (respectively, integrating respect to $1$) or summing up, if the indices are discrete, the distribution function $|\psi|^2(x_1,x_2)$ to obtain marginal distribution functions.

An {\it entangled pure state} is a pure state of a composed system which is not separable, namely, the classical principle of separability does not hold. In the quantum mechanical setting above, the pure entangled states are elements of the product Hilbert space $\psi\in\,\mathcal{H}=\,\mathcal{H}_1\otimes \mathcal{H}_2$ which are not product states. A simple case of entangled state is the {\it Bell' state}, which in Dirac's notation is of the form\footnote{The Einstein Podolski Rosen paradox discussed in terms of $1/2$-spin states appeared first in D. Bohm's book \cite{Bohm1951}, section 22.17.},
\begin{align*}
|spin=0\rangle=\,\sqrt\frac{1}{2}\left(|\uparrow\rangle_1\otimes \,|\downarrow\rangle_2-\,|\downarrow\rangle_1\otimes \,|\uparrow\rangle_2 \right),
\end{align*}
where $1$ and $2$ refers to two separate regions of spacetime.  The labels $\uparrow$, $\downarrow$ refers to the possible states of the spin at $1$ and $2$ along a given space direction. A Bell' state describes a system composed by two quantum particles with zero spin. Bell' states are of fundamental relevance for applications and in the theory of quantum entanglement.  Also, Bell's states are of relevance in the investigation of quantum non-locality. Note that, although spacelike separation  or timelike separation between $1$ and $2$ is not fundamental in the construction of Bell' states. However, the paradoxical conclusions of the analysis of this system arise when $1$ and $2$ are spacelike separated \cite{EinsteinPodolskiRosen}. In that case,  according with the theory of special relativity, it is not possible to send physical signals between the system at $1$ and the system at $2$, but, according to quantum mechanics, correlations between $1$ and $2$ occur \cite{EinsteinPodolskiRosen,Bell1964}.

For pure states, as for instance Bell' states and related states, there are several criteria and measures of entanglement. One of them builds on the {\it Schmidt's decomposition} in finite dimensional Hilbert spaces in terms of maximally projected product states. This is a type of decomposition in terms of product states. If the state is separable, then the Schmidt's decomposition is trivial, since coincides with the state itself.  Non-trivial decompositions reveal entanglement. If the state is not separable, there are several elements that appear in the Schmidt's decomposition, with weights less than $1$. The state is entangled iff the maximal projection coefficient of the Schmidt's decomposition is less than $1$ \cite{GarrisonChiao}.

\subsection{Entanglement in mixed states}
Mixed states describe physical systems where the a priori knowledge about the state is not maximal. Mixed states are usually used to describe ensembles of individual systems where there is a classical probability $p_i$ for the possible state $\psi_i$.  For mixed states, described by density matrices, the concept of entangled states is reduced to the concept of entanglement in the case of pure states. In particular,  an entangled mixed state is such that the ensemble $\{(\psi_i,p_i)\}$ that determines the statistical ensemble of the mixture contains at least one entangled state.

For mixed states, it is more difficult to provide a measurement of entanglement in terms of measurement of correlations, since the correlations due to the mixing and the quantum correlations are undistinguishable. That is, it is not possible to specify if an experimental correlation between observables is due to quantum fluctuations or statistical fluctuation of the mixture (see for instance \cite{GarrisonChiao}, section 6.4). A necessary condition for separability is the condition of {\it positive partial transpose of the density matrix} \cite{Peres1996}. For identical systems with $\mathcal{H}_1\cong\mathcal{H}_2$, the positive partial transpose is discussed in the Glossary.

\subsection{Entanglement of states described by continuous variables}
 A class of entanglement of particular relevance for quantum illumination and quantum radar protocols is the class of entangled states parameterized by continuous variable systems \cite{AdessoIlluminati2007}.
  A relevant  first example of this type of entanglement is realized by EPR-states of the form
\begin{align}
\psi(x_1,x_2)=\,\int^{+\infty}_{-\infty}\,\frac{dk}{2\pi}\,F(k)\,\exp \left(\imath\,k\,(x_1-x_2)\right).
\label{EinsteinPodolskiRosen states}
\end{align}
These states are null eigenvectors of the total linear momentum operator,
\begin{align*}
\left(\hat{p}_1+\,\hat{p}_2\right) \psi(x_1,x_2)=\,0,
\end{align*}
independently if the spacetime points $x_1,x_2$ are spacelike, timelike or lightlike separated. This is the basis for the Einstein Podolsky Rosen paradox.
 This particular class of states provides an example of quantum entanglement, namely, the anti-correlation among the linear moment of the two systems $1$ and $2$.

 These states are, un-physical, un-normalizable, but can be approximated by two-mode squeeze states. Thus in quantum mechanics, a continuous variable system with $N$ modes is described by the cartesian product of $N$ Fock spaces. Hence these systems are used to describe states of the electromagnetic quantum field. For a exposition of these theory, see \cite{AdessoIlluminati2007}. For a recent exposition of the theory of  Gaussian states and applications in quantum information, see \cite{Weedbrook}.
\bigskip
\\
{\bf Continuous mixed entangled states}.
For applications in quantum illumination and quantum radar, it is important to have a separability criteria for continuous entangled states applicable to mixed states, since several relevant protocols are based on continuous states. Several results have been obtained, among them the Peres-Horodecki partial transpose criteria for gaussian states discussed in \cite{Simon1999} and  the necessary and sufficient criteria for non-separability  discussed in \cite{Duan et al.}, which are applied in different protocols of quantum illumination.

\subsection{Discrimination of quantum operations and detection of targets}
The main difficulty in the applications of entanglement to enhance sensitivity in quantum measurements is that entanglement is a property easily spoiled by interaction of the quantum system with the environment. Such process of loss of quantum entanglement and in general, quantum coherence (quantum superpositions), is called decoherence. Very surprisingly, even if entanglement is lost, some of the protocols for quantum radar make use of a property of entangled states: entangled states used as a signal/ancilla system may imply an enhancement of entanglement-breaking channels discrimination \cite{Sacchi2005a,Sacchi2005b}. Quantum illumination protocols are fundamentally built on several realizations of this property.

Let us consider two quantum channels, represented by the operators $\mathcal{C}_1$ and $\mathcal{C}_2$. If $\rho$ is the initial state matrix density, the problem is to find the state $\rho$ such that the probability of discrimination for the output states $\mathcal{C}_1(\rho)$ and $\mathcal{C}_2(\rho)$ provides the minimal error.  In the case of bipartite entangled states, the initial state is a mixed state whose associated pure states are in a product space $\mathcal{H}\otimes\,\mathcal{K}$. If one of the parts is transmitted (signal) and the other is retained (idler), then the entanglement breaking channels produce outputs of the form $(\mathcal{C}_1\otimes I_\mathcal{K})\,[\rho]$ and $(\mathcal{C}_2\otimes I_\mathcal{K})[\rho]$ respectively, where $I_\mathcal{K}$ is the identity on the Hilbert space $\mathcal{K}$.

In the space of finite rank operators on $\mathcal{H}\otimes \mathcal{K}$ (operators whose trace is well defined and finite; see for instance \cite{Holevo 2011}, section 2.7), the trace norm is defined by  the expression $\|A\|_1 :=\,Tr\sqrt{A^\dag\,A}$. If the channel $1$ has assigned a probability $p_1$ and the channel $2$ has assigned a probability $p_2$, then for non-entangled states, the minimal error probability is of the form
\begin{align}
p'_E=\,\frac{1}{2}\,\left(1-\max_{\rho\in\,\mathcal{H}}\,\|p_1\,\mathcal{C}_1(\rho)-\,p_2\,\mathcal{C}_2(\rho)\|_1\right).
\label{probability of discrimination using classical states}
\end{align}
Instead, if one uses quantum entangled states one has
\begin{align}
p_E=\,\frac{1}{2}\,\left(1-\max_{\rho\in\,\mathcal{H}\otimes \,\mathcal{K}}\,\|p_1\,(\mathcal{C}_1\otimes I_\mathcal{K})[\rho]-\,p_2\,(\mathcal{C}_2\otimes I_\mathcal{K})[\rho]\|_1\right).
\label{probability of discrimination using entangled states}
\end{align}
Convexity properties of the space of states $\rho$ constructed as mixed states from $\mathcal{H}$ and from $\mathcal{H}\otimes \mathcal{K}$, linearity and convexity property
\begin{align*}
\|a\,A+\,(1-a)\,B\|_1\leq \,a\,\|A\|_1+\,b\|B\|_1
\end{align*}
implies that the maximum probability for distinguish $\mathcal{C}_1$ and $\mathcal{C}_2$, that is, the maximum for the probabilities $p'_E$ and $p_E$, is achieved for pure states. For pure states, one has that $p'_E\leq p_E$, that is, entanglement enhance discrimination sensitivity in entanglement-breaking channels. This is a counter-intuitive result, since the space $\mathcal{H}\otimes \,\mathcal{K}$ is larger than $\mathcal{H}$. Furthermore, the advantage of using entanglement is more evident for large dimensions of the Hilbert space \cite{Sacchi2005b}.
\\
{\bf Quantum entanglement or quantum correlations?}
The concept of quantum non-locality or quantum correlation differs from the notion of quantum entanglement. Indeed, particular quantum systems can show correlations beyond the ones allowed by classical distributions even in the case where there is no entanglement present \cite{Bennett et al.}. It is in this context that the notion of quantum discord is, besides the discussion on quantum channel discrimination, of interest in quantifying quantum correlations in a more general context than entanglement \cite{HendersonVedral, Ollivier Zurek}. The application of quantum discord as explanation of the advantages of quantum correlations respect to classical correlations in quantum illumination, even if entanglement has degraded, can be found in \cite{Weedbrook}.

\subsection{General approaches to quantum radar}
Quantum radar is the application of entangled light for radar use. They are type 3 and type 4 quantum sensors, according to the above classification. Indeed, different possibilities to use quantum entangled light for Several approaches  have been proposed in the literature:
\begin{itemize}
\item {\bf Interferometric quantum radar}. In this protocol, two entangled beams, the {\it idler}, which is retained, and a {\it signal} beam, which is sent to explore a particular region of interest.  After the signal beam probes the region of interest, there is a joint measurement performed at the receiver location where the phase difference between the two interfered beams is measured. Quantum interference with beams composed by entangled NOON-state allows to reach the Heisenberg limit in sensitivity in the difference in phase between the two paths followed by each beam, instead of reaching the standard quantum limit in the form of the shot noise limit as it appears in Mach-Zehnder interferometer \cite{Boto et al., Marco Lanzagorta 2011}. Other forms of photon beams, like those composed by photons in squeezed states, also beat the shot noise limit \cite{GiovannettiLloydMaccone2001, GiovannettiLloydMaccone}, even in some cases reaching the Heisenberg limit too.

    Interferometric quantum entanglement requires to keep the entanglement alive from the initial entangled beams through the whole process until the interferometric measurement of phase is performed. This  constraint of keeping alive the entanglement during the whole process limits the applicability of the scheme very much, due to the losses and attenuation \cite{GilbertHamrick,GilbertWeinstein2008}. The whole setting becomes very sensitive to noise.
\\
\item {\bf Quantum radar based on quantum illumination protocols}. The general scheme of  quantum illumination protocols is the following. An entangled source of light for signal beam with an idler beam are used to detection. THe signal is send to detect or track a possible space region where a target could be located; the idler is retained. The theory of quantum illumination  shows an enhancement in domain and sensitivity in several detection observables, for a theoretical low reflective, low signal intensity, high time-bandwidth product and bright noise background environment. These benefits are expect at optical frequencies \cite{Lloyd2008, Tan} and at microwave frequencies \cite{Barzanjeh et al.}. These gains in detection sensitivity are resilient to the loss of quantum entanglement during the round trip. Indeed, after entanglement is lost by the process of decoherence and attenuation, the residual correlations between the received beam (when the target is there) and idler beam   can be higher than for protocols working with entangled photon beams than for protocols based upon non-entangled light sources. A discussion of how quantum enhancement survives even when quantum entanglement has disappeared is provided in  Appendix \ref{AppendixLloyd quantum illumination}.
\\
\item {\bf Hybrid quantum radar systems}. These protocols are modified version of quantum illumination. In these protocols, entangled light is prepared to be used as a pair of signal/idler system as in a quantum radar illumination protocol. Again, the protocols rely on the enhancement of the residual correlations inherited from the original correlations due to the quantum entanglement. However, the detection and storage processes for both, the idler and the received beam, are achieved using  classical digitalization and storage methods. This protocols present advantages (it can capture and storage for long time information), but also introduces losses in sensitivity and enhancement respect to quantum radar based in full quantum illumination protocols. These hybrid methods have been discussed and experimentally demonstrated recently by several different groups \cite{Barzanjeh et al.2019, Luong et al. 2019}.
\\
\item {\bf Maccone-Ren theoretical quantum-radar protocol and further developments}. L. Maccone and C. Ren have developed a theoretical protocol for a quantum radar, able to determine the range and transverse displacement of a target. The fundamental idea is to exploit the advantage on the quantum sensing capability of EPR-type multiple entangled photons respect to illumination using non-entangled light states. In particular, it is shown that using beams composed by quantum states with $N$ entangled photons as signal beams, the standard error in the localization of the target is reduced in a factor $\sqrt{N}^3$ compared with the protocol of using $N$ non-entangled single photon states.

    Although Maccone-Ren's protocol effectively is a quantum radar, it presents special problematic issues. Among them is the sensitivity to environmental noise. We will discuss a combined  strategy between Maccone-Ren's protocol and quantum illumination to solve this problem and the so called range problem in quantum illumination all together, an idea further developed in \cite{Ricardo2020b}.

    \end{itemize}

While the first protocol works in a regime of entanglement, the other forms of quantum radar, in particular quantum illumination and hybrid quantum illumination and quantum illumination with multiple entangled photons, are applicable even under complete loss of entanglement. This makes such protocols particularly useful for radar applications, where entanglement is easily lost. But this potential application, to be physically realized, has to overcome several technical difficulties.
Indeed, all known quantum radar protocols present handicaps of theoretical and/or technological nature, that eventually lead to the loss partially or totally of the aforementioned benefits.

\section{Quantum Interferometric radar protocols}\label{quantum interferometric radar protocols}
The first type of quantum radars that we will consider are such they keep alive the entanglement during the whole detection process. They are called quantum interferometric radars. There are several comprehensive studies of them, among them \cite{Smith 2009} and more recently \cite{Marco Lanzagorta 2011}.
    \subsection{Theoretical protocols for quantum interferometric radar and experimental demonstrations}
Interferometric quantum radar is based upon the analogy of the general protocol of radar detection with Mach-Zender interferometry. In Mach-Zender interferometry with coherent light, a light beam passes through a beam splitter, that divides the beam in two beams and let pass each of the beams through different optical paths. Then the beams are recombined and an interference experiment measurement performed. The difference in the optical paths that the beams followed implies a difference on phase $\varphi$ between the quantum states when are re-joined. This can be easily seen via path integral interpretation of transition amplitudes as discussed in \cite{Sakurai}. Information on the characteristics of one of the paths can be extracted from $\varphi$. Information on  $\varphi$ can be obtained from the measurements of the intensity of the beams extracted from the receiver, that is, from the observed statistical distribution of individual photons arriving to the detector.

It turns out that when using coherent light as a source beam composed by $N$ independent photons, the error in the estimation of $\varphi$ grows asymptotically with the number of probes $N$ or with the number of photons of the beam, as $1/\sqrt{N}$. This is a direct consequence of the statistical independent character of the $N$ individual photons composing the light beam \cite{GiovannettiLloydMaccone} (see also the glossary in the appendix section).

The analogy between interferometric radar detection and Mach-Zender interferometry relies in the following protocol for the interferometric radar. Let us assume first that the source beam is coherent light. In the interferometric radar system, an initial source light beam is split in two beams. One of them is sent to explore an optical path (signal beam), while the second beam is keep alive in the detector during the whole experiment (idler beam). After the signal beam is received, it is recombined with the idler beam and an interferometric experiment (joint measurement) is performed. As for Mach-Zender interferometry, one can recover $\varphi$ from measuring intensities of the detected photon detection distributions.  Quantum metrology theory \cite{GiovannettiLloydMaccone2001, GiovannettiLloydMaccone} establishes that the error on $\varphi$ implies an error in the estimation of the target range of the form
\begin{align}
\delta R=\,\mathcal{O}\left(\frac{1}{\Delta \omega \sqrt{N}}\right),
\label{optical path sensitivity}
\end{align}
where $\Delta \omega$ is the signal band-width. An interferometric radar works as the case of a Mach-Zender interferometer, but where one of the arms of the interferometer (corresponding to the signal beam) is much larger than the other (see \cite{Marco Lanzagorta 2011}, section 5.2).

The analogy between quantum interferometry and radar sensing can be generalized to the case when beams are composed by entangled quantum states of light. $NOON$-states are among the type of quantum entangled states which are used in quantum sensing applications \cite{Boto et al.,Marco Lanzagorta 2011}. A NOON-quantum state is of the form
\begin{align*}
|\psi\rangle & =\,\frac{1}{\sqrt{2}}\,\left(\frac{(a^\dag_1)^N}{\sqrt{N !}}\otimes Id_2+\,Id_1\otimes \frac{(a^\dag_2)^N}{\sqrt{N !}}\right)|0\rangle_1|0\rangle_2\\
& \equiv  \,\frac{1}{\sqrt{2}}\left(|N0\rangle+\,|0N\rangle\right),
\end{align*}
where $|0\rangle_1|0\rangle_2$ is the vacuum state and $a_i,\,i=1,2$ are the annihilation operators for the photons associated to pass by arm $1$ or arm $2$. The effect of a phase shift $\varphi$ along one of the arms implies that the quantum state is of the form
\begin{align}
|\psi\rangle \equiv  \,\frac{1}{\sqrt{2}}\left(|N0\rangle+\,e^{\imath N\,\varphi}\,|0N\rangle\right)
\label{NOONstates}
\end{align}
It can be shown that the error in the estimation of the phase $\varphi$ when using a particular observable  for this $NOON$-estate   is of order  $\delta R \sim \,1/N$  \cite{GilbertWeinstein2008,Marco Lanzagorta 2011}. This implies the same order of precision for the estimation of the range $R$ of the target, that contrast with the asymptotic precision of order $1/\sqrt{N}$ when using coherent states. This precision in $1/N$ in the estimation range is generically known  as the {\it Heisenberg limit} \cite{MargulisLevitin}.  We also introduce the Heisenberg limit in the Glossary.
\subsection{Issues on the practical implementation of quantum interferometric radar}
The theory discussed above assumes ideal conditions. However, the Heisenberg limit cannot be reached using NOON-states in presence of attenuation due to absorption  and scattering of the beam with the propagation media \cite{Marco Lanzagorta 2011, GilbertWeinstein2008}. The effect of attenuation in the determination of range sensitivity has been discussed in \cite{Marco Lanzagorta 2011, GilbertWeinstein2008}, including the case of atmospheric attenuation \cite{GilbertHamrick}. In the model used in the literature, the quantum state in an attenuated media is of the form
\begin{align}
\nonumber |\psi\rangle & \equiv \, \frac{1}{\sqrt{2\,N!}}\,e^{-\left(\imath \, \eta_1\,\frac{\omega}{c}+\,\kappa_1(\omega)/2)\right)\,\left(N\,L_1\right)}\,\left(\hat{a}^\dag_1\right)^N\,|0\rangle_1|0\rangle_2\\
&  +\frac{1}{\sqrt{2\,N!}}\,e^{-\left(\imath\,\eta_2\, \frac{\omega}{c} +\,\kappa_2(\omega)/2\right)\,\left(N\,L_2\right)}\,\left(\hat{a}^\dag_2\right)^N\,|0\rangle_1|0\rangle_2,
\label{attenuate NOON state model}
\end{align}
where $\eta_i, i=1,2$ are refractive indices of the paths $i=1,2$, $\kappa_i(\omega)$ $i=1,2$ are attenuation indices, $\omega$ is the frequency of the radiation and $c$ is the speed of light in vacuum. The model implies for $L_2\approx\,l_1$ and $\kappa_1 <<\kappa_2$  an exponential attenuation along the path $2$ respect the path $1$,
\begin{align*}
|\psi\rangle \to\,\frac{1}{\sqrt{N!}}\,e^{-\left(\imath \, \eta_1\,\frac{\omega}{c}+\,\kappa_1(\omega)/2)\right)\,\left(N\,L_1\right)}\,\left(\hat{a}^\dag_1\right)^N\,|0\rangle_1|0\rangle_2+ \delta |\psi\rangle,
\end{align*}
where $\delta |\psi\rangle$ is an exponentially attenuated state respect to the first term.

It can be shown that as consequences of the attenuated state model \eqref{attenuate NOON state model} the following:
\begin{itemize}
\item Quantum interference using $NOON$-states does not surpass the standard quantum limit \eqref{optical path sensitivity} except in very low attenuation levels. Even it can perform worst than the standard quantum limit in realistic attenuation scenarios, a situation which is worsted as $N$ increases \cite{Gilbert Hamrick Weinstein2008}. This is a potential obstacle for the use of quantum interferometry using NOON-states for radar applications \cite{GilbertWeinstein2008}.

\item The effects in the error on the estimation of the range present periodic divergences in the azimuth angle \cite{Marco Lanzagorta 2011,Smith 2009}.
\end{itemize}
Note that the process of attenuation of propagation of photon states in atmospheric conditions is not fully identified with decoherence, but with a related dissipative process (see for instance \cite{Zurek2002} or the description of a model for decoherence in the Glossary in the Appendixes section). A complete treatment of atmospheric effects on quantum entanglement needs to consider decoherence effects.

\subsection{On the use of adaptive optics correction in quantum interferometric radar} In order to overcome the problem of the attenuation and the effect of the atmosphere on entanglement, it has been proposed the use of adaptive techniques \cite{Smith 2009, Marco Lanzagorta 2011}. The analysis presented in  \cite{Smith 2009} shows that a maximal range over $1000\, km$ can be achieved. Without adaptive techniques, a maximal range of $60\, km$ can be achieved. However, there is several idealizations on these estimates, as the use of ideal detectors.

The use of adaptive techniques requires the forehand knowledge of several relevant parameters of the target. Specifically, it requires the a priori knowledge of the target range. This limitation precludes the use quantum interferometric radar as a practical scheme for a realistic long range quantum radar. This problem will be indeed a recurrent issue in quantum radar protocols. Despite this,  the use of adaptive optics methods can imply benefits when using quantum interferometric radar as scanning systems.

The related but different effect of decoherence has not been considered in the analysis of J. Smith \cite{Smith 2009}. Indeed, quantum decoherence by interaction with a noise atmosphere environment is a threat for quantum interferometric radar that prevents to have the advantages of quantum entanglement in quantum technology.
\subsection{Other possible approaches to quantum interferometric radar}
Quantum interferometric radar could also be realized by means of  entangled states as described in \cite{Yurke et al. 1986,Dowling, GiovannettiLloydMaccone}. These states are of the form
\begin{align}
|\Psi\rangle =\,\frac{1}{2}\left(|N_+\rangle_A\,|N_-\rangle_B+\,|N_-\rangle_A\,|N_+\rangle_B\right),\quad N_{\pm}=\,\frac{N\pm 1}{2}.
\label{N+N- states}
\end{align}
 Similarly as for NOON-states, the error in the measurement of $\phi$ is of the form $1/N$, a reduction of the estimation error of order $\sqrt{N}$ respect to coherent light. The use of such states as models for quantum interferometric radar seems not been pursued literature.

 Another option to explode the enhancement of interferometry for radar applications could be the use of squeezed states that interfere with coherent states. In quantum metrology, squeezed state provide an enhancement that, although it does not reach the Heisenberg limit, still surpass the shot noise level \cite{Caves1981,BarnettFabre Maitre}. The use of squeezed states, by injecting them in port $1$ of the interferometer and a coherent state in the port 2, allows to achieve a sensitivity in the phase difference $\varphi$ of order $1/N^{3/4}$. This is an improvement compared with que standard quantum limit $1/\sqrt{N}$ that one reaches in the case the port $1$ is  injected a vacuum state \cite{BarnettFabre Maitre}. One can expect analogous results when squeezed states are applied in target detection.

\section{Quantum illumination} \label{Quantum Illumination}
The fundamental criteria for enhancement in sensitivity detection of quantum illumination protocols respect to the analogous classical illumination protocols are based upon the theory of quantum detection and estimation theory \cite{Helstrom1976}, which is a generalization of Chernov's theory \cite{Chernov}. The theory takes the ultimately form of quantum Chernov's theorem \cite{Audenaert et al.} in the case of a particular form of entanglement breaking channels.

In the particular case of  quantum illumination, the general setting is the following.  Two beams of entangled photons are generated.
 One of the generated beams will be used as a signal, while the other will be the idler, which is keep alive during the full experiment until measurements are done.
 The signal is sent to probe the region where the possible target could be and then it is detected back. The received beam is compared with the idler beam, typically by means of a joint measurement of the in-phase and quadrature voltages of the received and idler beams, but other forms of joint measurements are currently in use, as direct photo counting detection methods. Quantum illumination procedure leads to a theoretical enhancement  in sensitivity and/or signal to noise ratio respect to the use of classical light beams  with the same characteristics of brightness and energy. This happens in situations of low brightness signal, low reflectivity target and noise environment \cite{Lloyd2008, Tan, Barzanjeh et al.}.

The generation of the required entangled beams is usually achieved by means of spontaneous parametric down conversion methods \cite{Wong Shapiro Kim 2006, Caves Schumaker 1985} (in short SPDC or PDC) or variations of this mechanism, or other non-linear quantum processes generation. In SPDC generation, two beams composed by pairs of entangled photons at frequencies $\omega_s, \omega_i$ by pumping with a frequency $\omega_p$ a crystal with a second order non-linear susceptibility.
As a consequence of the method of production of the beams, in SPDC generation the idler and signal beams are correlated by the following characteristics  \cite{Lloyd2008, Marco Lanzagorta 2011, Shapiro2019}:
 \begin{itemize}
 \item Correlation in frequency. The idler and signal systems are correlated by conservation of energy, during the generation by the SPDC. The energy conservation implies the correlation for the frequencies of the photons
     \begin{align*}
     \omega_p=\,\omega_s+\,\omega_i,
     \end{align*}
 with $\omega_s$ the signal frequency and $\omega_i$ the idler frequency, implying energy at the photon level,
 \begin{align*}
 \hbar\omega_p=\,\hbar\omega_s+\,\hbar\omega_i,
 \end{align*}
 \item There is also conservation of momenta at the photon level,
 \begin{align*}
\hbar{\bf k}_p=\,\hbar{\bf k}_s+\,\hbar{\bf k}_i.
\end{align*}

  \item Correlation in arrival time. They are exactly correlated on the instant where both the idler and the signal are created by the SPDC process. Hence they must exactly meet in space and time when a joint measurement is performed. The correlation in time in the generation is at least up to $10^{-1}\,ps$ \cite{Burnham and Weinberg} and there is a delay between the beams due to differences in the refraction index in each optical path.

  \item The intensity of the idler and signal beams are the same, because for each of the photons in the idler beam, there is an entangled photon in the signal beam.
 \end{itemize}

Quantum illumination is resilient to noise and loss of entanglement due to decoherence \cite{Zurek2002}. This robustness is the main difference with respect to other quantum sensing protocols \cite{GiovannettiLloydMaccone} and makes it potentially useful in the development of radar and quantum sensing technologies. However, some non-idealities of the mechanism of detection,  combined with the form in which it is generated, implies that enhancement drop to lower figures respect to theoretical ideal performance. On the other hand, the enhancement is in first order of approximation independent from the loss of entanglement when the signal beam arrives at the receiver (see the discussion in Appendix \ref{AppendixLloyd quantum illumination}). This enhancement is due to the residual correlations due to the original entanglement with respect to the non-entangled sources is surprising, but it can be proved that this is the case by direct computation using techniques from quantum mechanics. Beyond computation, two ways to understand the enhancement is by means of its relation with the notion of entanglement enhancement sensitivity in entangled breaking channels \cite{Sacchi2005a,Sacchi2005b} and through the notion of quantum discord \cite{Weedbrook}.

In the following, we review several proposal for quantum illumination, illustrating the enhancement benefits and the problematic points that quantum illumination has from practical and theoretical point of views.

\subsection{Lloyd's proposal on quantum illumination}\label{Lloyd quantum illumination}Lloyd's original theoretical proposal compared the detection capability of two different optical transmitters. In the first, the beam is composed of un-entangled single-photon pulses. It is sent to probe a spatial region and received back after exploration. In the second, two entangled beams (the idler and the signal beam) are generated. The signal beam is sent to explore the region, while the idler is retained in the detector. After scattering with the target and being received the signal beam, the detector makes a joint measurement of the idler and received beams. Entanglement is assumed to be completely lost during the round trip.

The characteristics of Lloyd's quantum illumination protocol are the following:

\begin{itemize}

\item $N$ pulses with a single photon per pulse.

\item High time-bandwidth product $M=\,T W\,\gg 1$, where $W$ is the bandwidth and $T$ is the temporal detecting window.

\item Low reflectivity index $0<\eta\,\ll 1$ in presence of target. In absence of target, $\eta =0$.

\item The background light's average photon number per mode, $N_B$, satisfies the low-brightness condition $N_B\ll 1$.

\item For each transmitted signal pulse, at most one photon is detected by time, implying the condition $M\,N_B\ll 1$.

\end{itemize}
Under these assumptions, there are two different regimes of interest. For the single-photon state beam protocol, the {\it good regime} happens when $\eta/ N_B \,\gg 1$, while for beams formed by entangled photons, the {\it good regime} happens when $\eta > N_B/M$.  For non-entangled light, namely, single-photon states, the probability of error  $Pr^+(e)_{SP}$ is bounded as
\begin{align}
Pr^+(e)_{SP}=\,e^{-N\eta}/2,\quad \textrm{for}\, \eta\gg N_B,
\label{Lloyd good regime SP}
\end{align}
for single photon beams. For quantum illumination, the probability of error is bounded as
\begin{align}
Pr^+(e)_{QI}=\,e^{-N\eta}/2,\quad \textrm{for}\, \eta\gg N_B/M,
\label{Lloyd good regime QI}
\end{align}
showing an enhancement of the region of validity of this good regime (were the probability of false positive is very small) in the case of quantum illumination, despite that there is not an enhancement in the probability of false positive.

In the so called {\it bad regime} (the probability of error is in some sense large), we have the following probabilities of false detection,
\begin{align}
Pr^-(e)_{SP}=\,e^{-N\eta^2/8N_B}/2,\quad \textrm{for}\, \eta\ll N_B,
\label{Lloyd bad regime SP}
\end{align}
and
\begin{align}
Pr^-(e)_{QI}=\,e^{-N\eta^2 M/8N_B}/2,\quad \textrm{for}\, \eta\ll N_B/M.
\label{Lloyd bad regime QI}
\end{align}
Thus for very non-reflective systems, the probability of error in quantum illumination is reduced drastically with $M\gg 1$ respect to single photon states beams. Furthermore, there is an enhancement of the region of validity of this result, from $\eta\ll N_B$ to $\eta\ll N_B/M$.

In order to provide an example of  this advantage, let us consider an optical frequency of $300$ $THz$ and a pulse of 1 $\mu s$ with a $0.3$ percent of bandwidth yields $M\sim 10^6$. This reduces notably $Pr^-(e)_{QI}$ respect to $Pr^-(e)_{SP}$, that corresponds to $60 \, dB$ higher signal to noise ratio from quantum illumination respect to single photon illumination.

The condition of low-brightness background noise $N_B\ll 1$ is full-filled in the optical regime for the sky normal conditions \cite{Shapiro Guha Erkmen 2005}.
However, Lloyd's quantum illumination could be extended to bright backgrounds $N_B\gg 1$, as the second model considered by Lloyd itself suggested \cite{Lloyd2008}. Although, this is not a particularly physical condition at optical frequencies, since for sky day light in the optical regime $N_B\ll 1$, the condition $N_B\gg 1$ can be full-filled in presence of bright jamming.

Lloyd's analysis presupposes the following technical assumptions:
\begin{itemize}
\item There is a source of high tim-bandwidth product photon beams,

\item There is no losses in the storage system,

\item  The receiver  performs optimally: it detects individual pair of correlated photons.

\end{itemize}
Later developments have shown that such assumptions could be unrealistic and that non-ideal physical conditions can reduce considerably the theoretical benefits of quantum illumination. Specially difficult to implement are detectors with low or negligible losses.

\subsection{The work of Lloyd and Shapiro on coherent state and quantum illumination}
Despite the significant advantage of quantum illumination respect to the single-photon states illumination, coherent beams outperforms quantum illumination, as the analysis of Lloyd and Shapiro showed \cite{ShapiroLloyd}. Under the assumptions that
\begin{itemize}

\item An ideal laser produces $N$ pulses, each of which has unite average photon number,

\item For all reflectivity $0<\,\eta\leq 1$,

\item For low-brightness background, $1\gg N_B \geq 0$,
\end{itemize}
it can be proved that the probability of false positive for beams composed by coherent state is such that the Chernov's type bound
\begin{align}
Pr(e)_{CI}\,\leq e^{-N\eta}/2,\quad N_B<< \,1
\label{coherent illumination 1}
\end{align}
holds good.
Performance of coherent light equals to the performance of quantum illumination in the good regime, but outperforms when quantum illumination operates in the bad regime. The bound \eqref{coherent illumination 1} shoes that under the condition $N_B<<\,1$ there is only good regime. A way to understand this advantage of coherent $N$-modes with averaging number $1$ respect to single-photon quantum illumination with a single photon per pulse is to think that the property of coherence of the $N$-repeated transmissions of coherent states can be seen as a single coherent state with average number $\sqrt{N}$. For such modes, the error probability in of the form \eqref{coherent illumination 1} under the assumptions of the model, in particular, the assumptions $N\,\eta\, >>\, N_B$ and $N_B<<\,1$.

For the coherent state transmitter with homodyne detection, the error probability is of the form
\begin{align}
Pr(e)_{CI}\,\leq e^{-N\eta/2}/2,\quad N_B<< \,1.
\label{coherent illumination 3}
\end{align}
In the good regime, this is a $1/2$ factor in error exponential than quantum illumination \eqref{Lloyd good regime QI}, but in the bad regime it performs better than \eqref{Lloyd bad regime QI}.

The fact that an ideal coherent light illumination protocol theoretically outperform Lloyd's quantum illumination protocol in enhanced sensitivity and in signal to noise ratio could be seen as a strong limitation to the applications of quantum illumination, except in practical circumstances where the use of coherent light is disregarded.

 \subsection{Gaussian quantum illumination}
 The work from Lloyd and Shapiro \cite{ShapiroLloyd} showed that quantum illumination based upon Lloyd's protocol \cite{Lloyd2008} did not outperform a generic detection system based upon non-entangled single beams operating at the same energy and frequency characteristics, and indeed, quantum illumination could be substantial less sensitive than illumination protocols that employ coherent state light in situations of low bright noise environment.

 However,  quantum illumination stimulated further research on the potential enhancement on sensitivity when using quantum entangled states. Based upon the theoretical results for the theory of Gaussian states obtained and that allows to compute Chernov's bounds for discriminating in-equivalent unitary gaussian states  \cite {Pirandola Lloyd}, shortly after the work from Lloyd and Shapiro on coherent and quantum illumination Tan et al. \cite{Tan} showed that quantum illumination that uses Gaussian states theoretically outperforms any classical system, including coherent light illumination protocols. Apart from the gaussian nature of the quantum states of the light, the funcamental difference with previous protocols of quantum illumination is that Tan et al. protocol works better in bright noise scenarios\footnote{Although the original work of Lloyd considered the case of bright environment, the thermal bath model employed for his argument is of low practical relevance.}, where $N_B\gg 1$.

The theory of gaussian quantum illumination assumes the following conditions:
 \begin{itemize}

 \item The wave packets are composed by a number of photons per mode very small, $N_S\ll 1$.

\item High time-bandwidth product $M=\,T W\,\gg 1$.

\item Low reflectivity index $0<\eta\,\ll 1$ in presence of target. In absence of target, $\eta =0$.

\item The background light average photon number per mode, $N_B$, satisfies the high-brightness condition $N_B\gg 1$.

\end{itemize}
Note that the last condition in Lloyd's quantum illumination, namely, at most one photon is detected by time, implying the condition $M\,N_B\ll 1$, is dropped from the conditions in gaussian quantum illumination as developed in \cite{Tan}.

Under the above assumptions, quantum Chernov's bounds for the minimal possible error probability were obtained for the $H_0$ hypothesis (no object there) and the $H_1$ hypothesis (object there) for coherent light illumination and gaussian quantum illumination Tan et al. \cite{Tan}.
In that setting of quantum inference theory,  for the coherent light illumination it was found the relation
\begin{align}
Pr(e)_{CI}\leq \, e^{-M\eta N_S/4N_B}/2.
\label{coherent illumination 2}
\end{align}
In contrast, for gaussian quantum illumination it was found the relation
\begin{align}
Pr(e)_{QI}\leq \, e^{-M\eta N_S/N_B}/2,
\label{quantum illumination 2}
\end{align}
when $N_B\gg 1 $, $0<\eta \ll 1$ and $N_S\ll 1$. These bounds do not depend upon the type of detection (direct detection, homodyne detection or heterodyne detection). We remark that the bound found is the optimal case and valid for any possible detection scheme.

It is remarkable that in this regime, the bounds are very different than in the case of single-photon quantum illumination (compare the expressions \eqref{Lloyd bad regime QI} with \eqref{quantum illumination 2}). In particular, while the difference in the bad regime between single photon light and single photon quantum illumination was $M$ in the exponentials, in Gaussian quantum illumination is increased just by a fixed $4$ factor in the error probability exponential (or equivalently, 6 dB improvement) This improvement is independent of $M$, assumed large enough. This is a drastic reduction from the original assessment in Lloyd's analysis of quantum illumination, that claimed an improvement in sensitivity around 60 dB for $M\sim \,10^6$, but still concedes a theoretical ample advantage of quantum illumination respect to protocols based upon coherent state illumination.

It has been showed by Di Candia et al. \cite{Di Candia et al.} that $6\,dB$ in SNR is the theoretical maximal enhancement of quantum illumination respect to coherent light illumination, which corresponds in quantum communication setting, to use collective strategies (all modes $M$ are allowed to be measured altogether); when using local strategies (the modes are measured separatively, allowing for classical communication between them), the maximal  advantage is $3 \,dB$ in SNR. On the other hand, de Palma and Borregaard have shown that the two squeezed modes gaussian quantum illumination is optimal, in the sense that provide the maximal advantage respect to coherent quantum illumination \cite{de Palma Borregaard}.

\subsection{Receivers for Gaussian quantum illumination}
 It is not trivial to find a receiver for quantum illumination that can realize the theoretical advantages in SNR of quantum illumination for the following reasons. In Tan et al. theory the observable indicating the target presence is obtained from the observation of  the operator
 \begin{align*}
 \widehat{O}_{RI}(m):=\hat{a}_{Rm}\,\hat{a}_{Im}
 \end{align*}
  namely, the observable is the expectation value phase-sensitive cross-correlation $\langle \hat{a}_{Rm}\,\hat{a}_{Im}\rangle_{H_1}$
where
\begin{align*}
\hat{a}_{Rm}=\,\sqrt{\eta}\,\hat{a}_{Sm}+\sqrt{1-\eta}\,\hat{a}_{Bm}
\end{align*}
is the annihilation operators of the received mode $m$. $\hat{a}_{Im}$ is the annihilations operator of the idler mode $m$, $\hat{a}_{Bm}$ is the annihilation operator for the back-ground field and $\hat{a}_{Sm}$ is the corresponding operator for the signal mode. $H_0$ denotes the hypothesis that there is no target present, while $H_1$ is the hypothesis that there is target present.Theory gives the value
\begin{align*}
\langle \hat{a}_{Rm}\,\hat{a}_{Im}\rangle_{H_1}=\,\sqrt{\eta\,N_S\,(N_S+1)},\quad \langle \hat{a}_{Rm}\,\hat{a}_{Im}\rangle_{H_0}=\,0.
\end{align*}
For the classical light illumination, $\langle \hat{a}_{Rm}\,\hat{a}_{Im}\rangle_{H_1}=\,\sqrt{\eta}\,N_S$. Therefore, in the limit $N_S<<1$, the value of the quantum correlation exceeds the classical value of the cross correlation $\langle \hat{a}_{Rm}\,\hat{a}_{Im}\rangle_{H_1}$.
However, the operators  $\widehat{O}_{RI}(m)$ are impossible to be measured experimentally. The reason is that, although the operators $\widehat{O}_{RI}(m)$ can be expanded in individually observable quadrature operators, Heisenberg uncertainty relations precludes the simultaneous measurement and knowledge of all them four (two for the idler beam and two for the return beam).

An alternative detection procedure consists on measuring the phase insensitive cross correlations
$\langle \,\hat{a}^\dag_{Rm}\,\hat{a}_{Im}\,\rangle_{H_i}\quad\,\,H_i \,=0,1,$.
But for Tan et al. theory, this cross-correlation is zero is quantum illumination in both cases, when one assumes the  $H_0$ or when one assumes the $H_1$ hypothesis \cite{Tan}.

 Also, direct photo-counting is doubtless of help in quantum illumination. There is the believe that it cannot help to explode the theoretical advantages of quantum illumination\footnote{This view, that was keep at the time of these developments in 2009, should be revised under the light of current developments in photo-counting detection.}.
\bigskip
\\
{\bf Optical parametric amplifier detector}.
 The resolution of this implementation problem of Tan et al. theory required different reception methods.
A protocol based on this observable was developed by  Guha and Erkmen through his work on optical parameter amplifier (OPA) \cite{Guha Erkmen 2009}.
The theory of OPA is essentially based in a reverse type quantum optic process than spontaneous parametric down converted generation of entangled modes (SPDC). Both non-linear optical processes (see also the short introduction provided in Appendix \ref{non-linear optics section}). The OPA detector used for QI illumination uses single spatial mode fields where the signal mode with frequency $\omega_{S_0}+\delta\omega$ is only correlated with an idler mode frequency $\omega_{I_0}-\delta \omega$ such that both frequencies are in the phase-matching bandwidth condition $|\delta \omega|<\pi\,W$, being $W$ the bandwidth.

The observable in Guha Erkmen detector is based on expectation value of the number operator
\begin{align}
\hat{N}_T =\,\sum^M_{m=1}\,\hat{a}^{out\dag}_{Im}\,\hat{a}^{out}_{Im},
\label{Observable Guha Erkmen}
\end{align}
where $\hat{a}^{out}=\, \sqrt{G}\,\hat{a}_{Im}+\,\sqrt{G-1}\hat{a}^\dag_{Rm}$. The coefficient $G$ is the gain of the detector, that can be chosen to optimize the detection performance. $\langle \hat{N}_T \,\rangle_{H_i}$ determines the observable than can differentiate between absence and presence of target.  This observable can be expressed in terms of the correlations $\langle \,\hat{a}_{Rm}\,\hat{a}_{Im}\,\rangle_{H_i}$, $\langle \,\hat{a}_{Rm}\,\hat{a}^\dag_{Rm}\,\rangle_{H_i}$ and $\langle \,\hat{a}^\dag_{Im}\,\hat{a}_{Im}\,\rangle_{H_i}$. The main point is that now all these correlations are observable, since Heisenberg principle does not precludes measure them.

For OPA receiver in the usual regime of high bright environment and low signal intensity, low reflectivity and under the assumptions of lossless idler and perfect matching in time and phase between the returned and the idler beam, the Chernov's bound is of the form \cite{Guha Erkmen 2009}
\begin{align}
Pr(e)_QI\leq \,e^{-M\,\eta\,N_S/2N_B}/2 .
\label{Guha detector Chernov bound}
\end{align}

The OPA-reception protocol is not optimal and will not full-fill the theoretical $6\,dB$ advantage of quantum illumination over  coherent illumination, compared with the theoretical lower $3\,dB$ advantage in OPA detection.
\bigskip
\\
{\bf Sum frequency generation detector}.
 A proposal for an optimal receiver was discussed by Zhuang et al. in \cite{Zhuang Zhang Shapiro b}. They exploded the inverse process of SPDC, namely, sum frequency generation (SFG) process in the limit of large number of modes and low bright source. In continuous wave SPDC generation, a state of the form
 \begin{align*}
 \int^{\pi W}_{-\pi W}\,\frac{d\delta \omega}{2\pi}\,|\omega_S+\delta\omega\rangle\,|\omega_I-\delta\omega\rangle
 \end{align*}
 When a continuous generated SPDC idler-signal pair scatters  a second order crystal identical to the one used for quantum SPDC generation, with low probability, there is the possibility of the fusion of a idler-signal pair to generate a photon at the pump frequency. This happens for each of the pairs $(|\omega_S+\delta\omega\rangle,|\omega_I-\delta\omega\rangle)$ in the above integral i m a coherent way. This was used by Zhuang et al. to develop a receiver that under ideal conditions realized the $6 dB$ advantage from Tan et al.

SFG receiver does not have the problems that the original Tan et al. receiver and OPA non-optimum receiver has. Indeed, SFG receiver reaches the optimal Tan et al. bound in probability exponent \cite{Zhuang Zhang Shapiro b}. However, SFG receivers are based in several idealistic assumptions that are not reachable by current technology \cite{Shapiro2019}.

Zhuang et al. argued that a SFG receiver when conveniently complemented with a feed-forward mechanism, it is able to theoretical reach Tan et al. Chernov's bound \cite{Zhuang Zhang Shapiro b,Shapiro2019},
\begin{align*}
Pr(e)_QI\leq \,e^{-M\,\eta\,N_S/N_B}/2 .
\end{align*}

It is remarkable that the analysis of the FF-SFG receiver from \cite{Zhuang Zhang Shapiro b} allows to determine the receiver operating characteristic curve (ROC) curve. The ROC curve is the representation of the probability of detection respect to the probability of false alarm. Zhuang et al. showed an improvement respect to coherent state illumination \cite{Zhuang Zhang Shapiro c}. However, ROC curves provide more complete understanding of a receiver than SNR, since ROC curves are not based upon bayesian assumptions.

Despite its theoretical significance, due to current limitations in technology, the FF-SFG receiver is beyond a physical realization.

\subsection{Lopaeva et al. experiment}
Experiments demonstrating Lloyd's quantum illumination theory have been performed. The first one was carried out by Lopaeva et al. \cite{Lopaeva et al.}. In such experiment, the generation of the entangled states of photons was achieved by means of SPDC. The target was a highly reflective object located at a fixed, known position. The mechanism of SPDC generates a pair signal/idler beams  at optical wavelengths, appropriate  for gaussian quantum illumination. After scattering with the target, the signal beam was received in a highly efficient coincidence-counting receiver CDC camera, while the idler beam was also detected with the same type of CDC camera that the one  used for the detection of the signal beam. The classical illumination was generated by first stopping one of the beams generated in SPDC; the remained one was split in a signal and in an ancilla beam. In the experiment, the average number of photons per mode was $N_S\approx \,0.075$.

The results of the experiment show a clear enhancement of quantum illumination respect to classical illumination, given by the ratio between the signal to noise ratio of the quantum and classical light radars,
\begin{align}
QE=\,\frac{SNR_Q}{SNR_C}.
\end{align}
It was demonstrated in the experiment from Lopaeva et al. that the quantum enhancement $QE$ is larger than $1$. Indeed, for low intensity beams, with average number of photons per mode $N_S<<\,1$, the enhancement parameter can be of several orders of magnitude.

It is important to note that Lopaeva et al. experiment do not compare entangled light illumination with the benchmark classical illumination, namely, coherent light illumination. This point has been raised several times
in the literature \cite{Zhang et al. 2015, Shapiro2019}.

\subsection{England et al experiment and direct photon detection methods}
A  recent experiment on quantum illumination has been reported by England et al. \cite{England Balaji Sussman 20019}. The experiment compares the performance of an standard detector system that works using  non-entangled light with a system working with quantum illumination. The general idea of the experiment follows the previous experiment by Lopaeva et al. \cite{Lopaeva et al.}, but there are significant differences between both experiments.  In the quantum illumination experiment of England et al., the source of quantum entangled states is an spontaneous four wave mixing generator (SFWM). This procedure of generating entangled states  is based upon an effective four photon interaction \cite{Chen 2005,Smith et al.}, where the interaction of a powerful pump with a non-linear birefringent media produces two entangled photon state \cite{Smith et al.} (SPDC based on an effective three photon interaction \cite{Kwiat et al. 1995},\cite{GarrisonChiao}, section 13.3).
Also, the target used in the experiment is different than for Lopaeva et al. experiment, since in the experiment from England et al.   a diffusive target situated at a constant distance of $32 \,cm$ from the detector (and source) is investigated. For the quantum illumination scheme, two beams are generated. One is send to explore the presence of an object, while the second is retained in the detector system. As in Lopaeva et al. experiment, the classical illumination source is obtained by using the same signal beam as when the system operates in the quantum regime but disregarding the idler beam. The intensity of the classical light and quantum light illuminations are the same.  After the signal is received, there is a joint measurement of both beams.

 The detection system is based upon single photo counting receiver, in the case of classical illumination, and by the use of coincidence event receiver counting, in the case of quantum illumination. Therefore, the signal to noise ratio was defined by the phenomenological expression
\begin{align}
SNR=\,\frac{N_{in}-N_{out}}{N_{out}}
\end{align}
for both, the classical and the quantum illumination systems. In this expression,
$N_{in}$ is the number of detected photons where the target is there and $N_{out}$ is the number of photons where the target is not there.

The theory of spontaneous four wave mixing generation provides close expressions for the SNR in the case of classical and quantum illumination.
The generation of light by the SFWM mechanism has several interesting characteristic, detailed as follows:
\begin{itemize}
\item It can be proved that the enhancement, measured as the ratio between the quantum illumination $SNR_Q$ and the classical illumination $SNR_C$ is independent (in first approximation) of the losses in the detector \cite{Smith et al.} and also of the intensity the intensity of the laser jamming.

\item The non-classicality of the signal received is equivalent to the condition
 the bound
 \begin{align}
 QE=\,\frac{SNR_Q}{SNR_C}\geq 2
 \end{align}

\item Quantum enhancement is larger when the power of the signal beam is low, that is, when the average number of photons per mode satisfies $N_S<<1$. This enhancement is a distinctive character of the mechanism of generation SFWM of the pair of entangled photons.

\end{itemize}

In the experiment,
 the frequency of the jamming laser determines the time bin. Also, it is assumed that there is maximum of one photon per bin.
In the case of low noise environment, although there is a decrease respect to the theoretical gaussian quantum illumination $6\, dB$ benchmark, there is still benefit respect to classical illumination when the target is not there. However, this part of the experiment operates in the low-bright regime $N_B\ll 1$, a regime where quantum entangled states do not provide theoretical advantage over coherent light illumination. The experiment also shows that in such conditions classical illumination works well enough for detection.

The experiment also investigated quantum illumination in presence of jamming, showing a clear benefit of quantum illumination over classical illumination. This is shown by the signal to noise ratio, using coincidence detection (quantum illumination) or single detection (classical illumination). The experiment showed a clear benefit of the quantum case respect to the classical illumination.

As in Lopaeva et al.,  the experiment \cite{England Balaji Sussman 20019} did not compare quantum illumination with coherent light illumination. Hence it is open to the same criticism that usually is raised for Lopaeva et al. experiment.

\subsection{Experimental comparison of gaussian quantum illumination versus coherent illumination}
The experiment from Lopaeva et al. on quantum enhancement of quantum illumination respect to classical raised several criticisms. As it was discussed in  \cite{Zhang et al.}, Lopaeva et al. experiment compares a non-optimal source of classical light (thermal states) with quantum illumination. But  coherent states are the benchmark for classical illumination sensitivity, instead of thermal states. Furthermore, the CDC camera used in Lopaeva et al.  experiments are far from being the most efficient detection method, that it turns out to be homodyne-detection receiver for coherent light.

Using OPA method of detection, Zhang et al. reported the first experimental demonstration of enhancement of gaussian quantum illumination respect to coherent light homodyne-detection scheme was demonstrated \cite{Zhang et al. 2015}. In Zhang et al. experiment, a laser pump at $\lambda_p =\,780\, nm$ is used in a SPDC process to generate two entangled beams at wavelengths $\lambda_s =\,1590 \,nm$ and $\lambda_i =\,1530\, nm$. Further, noise is added at the same wavelength than the signal beam. The recombination of the idler and signal-noise beam are detected at a OPA detector (instead than a CDC-camera). In theory, the Guha et al. OPA detector can enhance up to $3\,dB$ in signal o noise ratio (SNR) when using quantum illumination respect to coherent light illumination detection. The reported improvement of quantum illumination respect to coherent light illumination is rather modest, of order 20\% in signal to noise ratio. The reduction on the efficiency between experiments and the theory elaborated by Tan et al. is explained by the multiple non-idealities of the experimental scheme and detector procedure.
\begin{comentario}
{\bf General remarks on experiments on quantum illumination}. Let us note the following:
\begin{itemize}
\item In all  experiments \cite{Lopaeva et al.,Zhang et al. 2015,England Balaji Sussman 20019}, the enhancement in signal to noise ratio is higher for low average number of photons per mode, when the condition $N_S<<1$ is meet.

\item In the above discussed experiments, except in the recent experiment of England et al.,  the target range must be known. Also, the target was fixed at a given range.

\item The design of the experiments of Lopaeva et al. \cite{Lopaeva et al.}, Zhang et al. \cite{Zhang et al. 2015} and England et al. \cite{England Balaji Sussman 20019} are on the optical wavelength, which is unpractical for long range radar applications, although it can potentially be applied as  non-intrusive short range quantum LIDAR or as a high sensitive, non-intrusive scanning applications.

\end{itemize}
\end{comentario}

\subsection{Impracticability of gaussian optical quantum illumination for radar applications}
Apart from the range problem discussed above,
the main problem that precludes the use of gaussian illumination for realistic radar applications in the optical regime is related with the conditions of brightness $N_B \gg 1 $ that gaussian quantum illumination assumes. In the optical regime (were sources of entangled beams by means of parametric down converted methods are relatively easy to find) do not hold for normal sky light conditions, where $N_B(optical sky) \ll 1$. Henceforth one arrives to the conclusion that for long range radar purposes, gaussian quantum illumination in the optical regime will be impractical, except in contra-jamming measures applications.

\subsection{Microwave quantum illumination}
 The stronger limitation of optical gaussian quantum illumination  for radar applications relies  on the inadequate implementation under which optical quantum illumination works for daily sky normal conditions. This problem is overcome in microwave quantum illumination by the use of microwave gaussian beams as signals. At microwave wavelength, daily sky conditions implies $N_B\gg 1$. It is a high noisy environment, which is one of the premisses to exploit the benefits of gaussian quantum illumination respect to the classical benchmark coherent light illumination, as it was experimentally demonstrated in \cite{Lopaeva et al.,England Balaji Sussman 20019}.

In this context, the proposal for quantum illumination in the microwave regime of Barzanjeh et al. \cite{Barzanjeh et al.} from 2015 was the following. Two beams of gaussian  entangled photons states are generated at optical frequencies using a SPDC system. The two beams have the same intensity. One of the beams is retained in the receiver as the idler and keep alive for a later joint measurement; the twin beam pass through an electro-optomechanical (EOM) cavity that converts the optical to microwave signal. After collimation, the microwave signal is sent to probe the region and later it is detected and converted back to another electro-optomechanical converter, that converts back from optical to microwave signal. After this, a joint measurement is performed of the detected/converted signal and the retained idler beam as described by Tan et al. gaussian quantum illumination theory. The detection method is heterodyne detection.

The theoretical probability of error  $Pr(e)_{QI}$ in the operating regime of gaussian quantum illumination $(0< \eta\ll 1, \,N_S\ll 1,\,N_B\gg 1)$  depends upon the characteristics  of receiver, implying an overall gain respect to coherent illumination that range up to 6 dB advantage in error probability exponential \cite{Tan}, by using the FF-SFG receiver of \cite{Zhuang Zhang Shapiro b}. The original analysis employed  receivers with theoretical gains in $Pr(e)_{QI}$ equivalent to 3 dB \cite{Barzanjeh et al.}.

There are additional losses at the receiver. To retain the idler beam using optical storage methods implies the generation of noise at a rate around $0.2 \,dB/km$ (noise induced/fiber propagation). This implies that, in order to keep some of the theoretical advantage from the original $ 3\,dB$, the range of the maximal radar range should be restricted to approximately $11$ km \cite{Barzanjeh et al.}.

One possible to treat the problem of the storage of the information coming the idler mode that has been initially discussed in the literature \cite{Barzanjeh et al.} is the use of {\it quantum memories} \cite{Simon et al.,Zurek2002}. Although the implementation of quantum memory technology for quantum illumination is technologically demanding and still to be achieved, it could potentially solve the storage problem in microwave quantum illumination, allowing to pass from  $3 $ to $ 6$ dB theoretical increase performance for longer range detection, as indicated initially in \cite{Barzanjeh et al.}.
\begin{comentario}
The method proposed by Barzanjeh et al. posses the following problem of the significant the loses in the optical to microwave signal conversion. Current conversion efficiencies are low (around $5 \%$), which is a problem to keep intensities large enough for radar applications.
\end{comentario}

\section{Issues in the implementation of quantum illumination for radar applications}\label{problems of the microwave qi}
It is clear by now that several difficulties are present at the experimental level in the form of non-optimal technological implementation and also at the fundamental level of the quantum illumination protocol for its implementation in radar applications (see for instance \cite{Pirandola et al. 2018, Shapiro2019}). In this section we review several of such specific issues.

\subsection{Reception and process of the correlated signal/idler beams}
Quantum illumination requires the identification of the correlated photons of the signal received beam and idler beam at the moment of detection. For several relevant receivers, like OPA and FF-SFG, this requirement implies prior knowledge of the target range \cite{Lloyd2008, Marco Lanzagorta 2011, Chang et al. 2019, Luong et al. 2019, Barzanjeh et al.2019}. Because of this constrain, it has been suggested the use of quantum illumination in applications where the target range is known by other means, as a scanner system or in bio-medical applications \cite{Marco Lanzagorta 2011, Barzanjeh et al.2019, Karsa et al.}, instead than as a protocol for a complete radar system. The advantage of quantum illumination over classical illumination can be exploded to have better resolution of the structure of possible targets, once the target has been located by other means. Note that this problem, which is the range problem, is a generic feature of quantum illumination.

There are other theoretical protocols for quantum radar that provide theoretical solutions to this problem   (see below Sections \ref{Maccone Ren protocol}, \ref{quantum illumination with multiple entangled photons}, \ref{Other protocols}). Indeed, the methodology followed by England et al. experiments \cite{England Balaji Sussman 20019} provides a simple method to determine the target in the case of low bright environment by exploding the coincidence detection method is similar to the methodology discussed in \ref{quantum illumination with multiple entangled photons}. However, each of the proposed solutions comes with new disadvantages, as we will discuss later sections.

\subsection{Generation of the entangled beams at microwave frequencies}
 The standard protocols for quantum illumination require the generation of quantum states composed by entangled pairs of photons. In the optical regime, the generation of the states is usually achieved by the use of SPDC techniques \cite{Kwiat et al. 1995}. However,
one problem to implement quantum illumination for radar applications  is the generation of entangled states within the frequency regime adequate to radar applications, namely, signal beams in the microwave frequency.
 Two different techniques has been used that remain fidel to the original quantum illumination/radar protocol:
\begin{enumerate}
\item {\bf Frequency conversion}.  The use of electro-optomechanical converters to pass from optical frequencies to microwave frequencies and viceversa, in the generation of the signal beam and in the detection of the reflected beam, respectively \cite{Barzanjeh et al.}. It was demonstrated using this method the reliability of microwave quantum illumination at the laboratory level (under cryogenic conditions).

A major deficiency of this technique of microwave entangled pairs generation is the low rate of the conversion from optical to microwave frequencies and viceversa. Intense sources of optical SPDC generation are required.
This is also true for alternative form of generation, as four wave mixing generation, specially if the applications are for long range sensing.

\item {\bf Josephson amplification}. Non-degenerate Josephson parameter amplifier (JPA)  \cite{Yurke et al. 1989, Lahteenmaki, Chang et al.2018, Barzanjeh et al.2019} has been used as a quantum microwave source in quantum illumination and quantum radar \cite{Chang et al. 2019, Luong et al. 2019,  Barzanjeh et al.2019}. JPA generators have been applied  in the new proposal of {\it quantum-enhanced noise radar} \cite{Chang et al. 2019} and in the experimental verification of the advantages of quantum microwave illumination at macroscopic distances and room temperature \cite{Barzanjeh et al.2019}.

The weakness of these methods is the extremely low temperatures that the JPA requires for operation. Also, the low intensity of the beams achieves precludes the systematic use of JPA microwave quantum illumination in long range applications.
\end{enumerate}

It is clear that the problem of generating entangled signal/idler states of pairs of photon beams in the  microwave regime suitable for radar operation remains an important factor for the further development of quantum radar. Indeed, to have other forms of generation with potential applications to quantum radar is one of the most relevant research lines in the fields. Let us mention that direct generation without the use of cryogenic techniques

\subsection{Losses introduced by the storage of the idler beam} One source of strong limitation for the application of quantum illumination to long range radar is the storage of the idler signal. For short range radar applications, for long range applications, the idler in l losses possess a limit on the maximal target range where quantum illumination implies an effective advantage in error probability exponent over coherent light illumination and in general, classical radar protocols.
If the idler transitivity is $\eta_I$, with $0<\eta_I\leq\,1$, then the Chernov's bounds are modified to the form \cite{Shapiro2019}
\begin{align}
Pr(e)_QI\leq \,e^{-M\,\eta\,\eta_I\,N_S/2N_B}/2
\label{OPA detector with idler loses}
\end{align}
for OPA detector and
\begin{align}
Pr(e)_QI\leq \,e^{-M\,\eta\,\eta_I\,N_S/N_B}/2
\end{align}
for SFG and FF-SFG receivers.

The use of optical fiber delay line implies a limitation of the range to a maximum of approximately 11 km, assuming a fiber loss of 0.2 dB/km and fiber propagation speed around 2/3 c \cite{Barzanjeh et al.}.

 The proposal of using a quantum memory for storage of the idler beam could potentially improve this range limit, since the efficiency in some quantum memories reach up to order 89\% \cite{Heshami2016}.

 Other methods to solve the loses in the idler storage apply classical digital storage of the idler and matching filter techniques. These methods will be discussed in later sections.
 \\
{\bf Losses due to atmospheric absorbtion}. This is another important source for losses. This difficulty has been already discussed before when we discussed quantum interferometric radar. This problem could have a similar treatment by means of adaptive optics correction \cite{Smith 2009}, with the same limitations and constrains in practical implementation.

\subsection{The time-bandwidth problem} Hight time-bandwidth product is essential for the theoretical advantage of quantum illumination respect to classical illumination. But in the microwave regime this is difficult to achieve \cite{Pirandola et al. 2018, Shapiro2019}. The mechanisms proposed in \cite{Barzanjeh et al.} of optical conversion to microwave implies a narrow band generation. Even using broadband amplifiers \cite{Macklin et al.}, the time-bandwidth product is very small in the microwave compared with what it could be easily available at optical frequencies. For example, a $1/3$-percent of fractional bandwidth at $1$\,$ \mu m$\,wavelength that provides $1 \mu s$ pulse duration at $10^6$ time-bandwidth product, only gives $10^2$ bandwidth at $1 \,cm$ wavelength.

Pushing to the mm-wave operation these figures will imply a decrease the efficiency on the single pulse bin interrogation efficiency of quantum illumination, spoiling the eventual advantage \cite{Shapiro2019, Pirandola et al. 2018}.
\subsection{Rayleigh-fading targets} Rayleigh-fading distributed amplitudes arise in situations when the surface of the target are not sufficiently smooth, in particular, at lidar amplitudes. In such more realistic conditions, the return signal is speckled, that is, the amplitude  is Rayleigh distributed and the phase is uniformly distributed. Indeed, for the amplitude the probability distribution at lidar amplitudes is a of the form $f(\delta \eta)=\,\frac{2\,\delta\eta}{\bar{\eta}}\,\exp (-\delta \eta^2/\bar{\eta})$ for $x>\,0$,  where here $\eta$ is a measure of the amplitude (reflectivity), $\delta \eta$ is the displacement respect to the average $\bar{\eta}$; for the phase, the distribution is of the form $f(\phi)=1/2\pi$. Tan et al. theory did not consider this effect, assuming known amplitude and phase of the return light. In fact, Rayleigh-fading spoils the $3\,dB$ advantage of OPA detectors \cite{ZhuangZhangShapiro}. On the other hand, it was also shown in \cite{ZhuangZhangShapiro} that when working with the FF-SFG receiver, quantum illumination outperforms classical light illumination benchmark, namely, coherent light illumination.

In any case, the Rayleigh fading effects reduces the benefits of quantum illumination over coherent light illumination, being the advantage much smaller than in the case of smooth targets. Also, the analysis done in \cite{ZhuangZhangShapiro} assumes a near loss-less idler storage and near unity efficient of the SFG receiver.

Let us mention, however, that the fadding problem applies to lidar frequencies and is not so relevant at microwave frequencies.

\section{Hybrid quantum illumination protocols}\label{Hybrid protocols}
Hybrid quantum radar protocols described below employ entangled light for the protocols of the illumination, but the mechanisms for the reception the signal beam are based on digital classical methods and classical matching filter techniques. In this section, we will describe the research work of two groups where these new proposal methods have been investigated and experimentally demonstrated \cite{Luong et al. 2019,Barzanjeh et al.2019}. For both, the following guidelines describe the corresponding protocols, that we present in closer form as exposed in \cite{Luong et al. 2019}.

\subsection{Quantum radar prototype of Chang et. al. and experiments}

The radar prototype described in  \cite{Chang et al. 2019}-\cite{Luong et al. 2019} develops and demonstrates experimentally the concept of quantum-enhanced noise radar (also named {\it quantum two-mode squeezing radar} in \cite{Luong et al. 2019}). It is a demonstration of how the use of quantum entangled signals/idler beams can outperform a classical equivalent two mode noise radar system working under the same conditions.

The radar experiment performed has the following steps \cite{Luong et al. 2019}. For entangled light illumination system, the protocol is the following:
\begin{enumerate}
\item Generation of two correlated noise beams at the JPA and their amplification (out of the JPA).

\item One of the signals (the idler beam) is amplified and measured immediately after generation using classical digital techniques. The result is stored using classical digital techniques. In particular, the in-phase and quadrature voltages for the idler signal are measured and recorded.

\item The  signal beam is amplified and sent to explore a spacetime region where the target could be located. In the experiment, the signals are sent through free space.

\item Receive and measure the signal using classical digital techniques like it was previously done for the idler beam.

\item Declare a detection if the detector output, after the application of matched filtering techniques, exceeds a given threshold value.
\end{enumerate}

The methodology is analogous for the case when the signal/idler beams are not entangled (two mode noise radar, in short TMN radar) than for the case entangled generated sources (quantum two mode squeezing radar, in short QTMS radar), making the standards for the protocol comparison the same.
 However, The  difference between the TMN radar and the QTMS radar arises in the generation of the corresponding sources.

For the QTMS radar, the sources are generated by means of a  Josephson parametric amplifier (JPA), that generates two entangled photons at microwave frequencies $\omega_1 =\,7.5376\, GHz$ (corresponding to a wave length of approximately $43\, cm$), $\omega_2=\,6.1445\, GHz$ (corresponding to a wave length of approximately $18\, cm$). After the generation, the signals are amplified.
Note that the amplification process introduces noise in the system and also is a cause of the lose of entanglement\footnote{The notion of entangled sources used in \cite{Luong et al. 2019} is a pragmatical one: they say that two quantum systems are entangled if the corresponding correlation matrix show higher correlations than the corresponding correlation matrix for classically correlated beams operating at the same specific characteristics. This is a pragmatical notion of entanglement, different from the notion of entanglement associated with the fact that the pair of photons generated at the JPA are entangled in a quantum mechanical sense. Indeed, the notion  in Luong et al. of quantum entanglement is referred as  non-classical or quantum correlation in quantum mechanical literature.}.  The idler (with $\omega_2=\,6.1445\, GHz$) is digitalized and stored as a classical record immediately after generation. The signal coming directly from the JPA is $-145.43\, dBm$, after which it suffers an amplification to the power $-82\, dBm$. The amplified signal (with $\omega_1 =\,7.5376\, GHz$)  is sent to probe the spacetime region.
The received signal is measured using heterodyne methods, digitalized after being received. Both are stored and compared as classical records using filtering techniques.

For the TMN radar, the radar prototype has the following characteristics. The generation of the signal is as follows. A carrier signal at $6.84105 \,GHz $ is generated and mixed with Gaussian noise centered at $069655\, GHz$ and band-limited width of $5\, MHz$. This will produce two sidebands at frequencies $\omega_1 =\,7.5376\, GHz$, $\omega_2=\,6.1445\, GHz$, as in the case of $QTMS$ radar protocol. The signal beam is then treated through the same amplifiers chain than for the QTMS radar signals. After the signals passes through an splitter, the signal $\omega_2$ is then detected by heterodyne methods and the results digitalized, while the signal $\omega_1$ reaches an X-band antenna horn and is then send it through free space.

Further specification of the experiment are the following.
The total power injected in the transmit horn was $-63\,dB$;
The power of the Gaussian noise generator, after discounting the noise of the amplification is $-82\, dBm$. Hence the signal to noise ratio is of order $-19\,dB$.
The implementation of the above procedures ensures that the QTMS radar and the TMN radar prototypes operate under the same conditions and characteristics.

The reception mechanism consists of the same kind of $X$-band horn antenna as in the case of QTMS radar connected to an amplifier at $25\, dB$. Then the amplified signal is connected to a digitizer performing measurements at $\omega_1=\,7.5376\, GHz$.

The free space that separates the two horns is $R=\,0.5 m$. In the experiment, there was no target, only free separation space between antenna horns.

\subsection{Covariant matrix for classical correlated noise radar and quantum correlated noise radar}
The detector functions that Luong et al. considered are constructed from the in-phase and quadrature voltages of the signals $1$ and $2$, where the signal $1$ is the one sent to probe the region of the space; the signal $2$ is measured right after generation and stored using conventional methods. Let us consider the four dimensional vector
\begin{align*}
x^\top:=(I_1, Q_1, I_2, Q_2).
\end{align*}
When the two signals are generated, the covariant matrix has the form \cite{Luong et al. 2019}
\begin{align}
E[x\,x^\top](0)=\,
\begin{bmatrix}
R_{11} & R_{12}(0) \\
R_{21}(0) & R_{22}
\end{bmatrix}
\end{align}
where each $R_{ij}$ is a $2\times 2$ matrix. Assuming stationary signals, the block matrices $R_{11}$ and $R_{22}$ do not depend upon time. After a time of evolution, the covariant matrix will be of the form
\begin{align}
E[x\,x^\top](t)=\,
\begin{bmatrix}
R_{11} & R_{12}(t) \\
R_{21}(t) & R_{22}.
\end{bmatrix}
\end{align}
It is on the off-diagonal blocks $R_{12}(t)$ and $R_{21}(t)$ where the information on the absence or presence of the target is encoded. The cross-correlations are characterized by two parameters, $\rho$ and $\phi$. $\rho=0$ characterizes the case when there is no target  and all the signal detected in the second antenna is due to envirommental thermal noise; the case $\rho\neq 0$ characterizes the case when there is a target. In the experiment the target presence is equivalent to apply a non-zero radar signal, either for the QTMS radar or for the TMN radarar.
Further details of this formalism will be introduced later when dealing with Shapiro's argument on asymmetric idler/signal analysis of coherent illumination.
\\
{\bf Experimental demonstration}
Experimental demonstration of the above quantum radar prototype has been performed. Instead of evaluating the signal to noise ratio, the receiver operating characteristics curves (ROC curves)\footnote{The precise way the probability of false positive and probability of detection are analyzed can be found in \cite{Luong et al. 2020}.} are plotted and analyzed for different experiments with different number of photon pairs detected. In the regime when the probability of false alarm is very small (the typical figure of order 0.001 or lower), there  is a remarkable factor of $10$ enhancement in the probability of detection respect to TMN adar protocol based upon classical correlated sources \cite{Chang et al. 2019, Luong et al. 2019}. Another advantaged show by the experiments is that the required integration time of the QTMS radar can be reduced in a factor up to eight respect to the operation time of the TMN radar.

By the nature of the technique used, further degradation of the theoretical maximal enhancement of quantum illumination respect to classical illumination of $6\,d B$ \cite{Tan} is introduced. Still, it is observed an improvement of QTMS radar  respect to the TMN radar when comparing the corresponding

Note that although in these experiments the JPA was settle in a cryogenic environment, the target or the trip of the signal beam can be located at room temperature, without target. On the other hand, the JPA itself is very sensitive to noise. It must be kept at very low temperature of $7\, mK$, in order to produce a vacuum respect to frequencies above $4 \, GHz$.

\subsection{Quantum radar prototype of Barzahjeh et al. and experiments}
The general concept and techniques behind the prototype of quantum radar discussed in \cite{Barzanjeh et al.2019} are similar to the QTMS radar discussed in \cite{Luong et al. 2019}, specially in the generation of the quantum entangled source and in the use of classical digital techniques in the detection process.

 In the work of Barzanjeh et al. \cite{Barzanjeh et al.2019} quantum illumination is also compared  against classically correlated coherent state illumination with the same specifications and prepared under the same characteristics.  In the experiment performed by Barzanjeh et al. there is  target in space (the target was absent in the experiments of Luong et al.\cite{Luong et al. 2019}).

For the quantum illumination, the protocol followed by Barzanjeh et al. had the following steps:
\begin{enumerate}
\item Two entangled microwave beams are generated directly from a JPA source.

\item The idler beam is measured using heterodyne detection and recorded right after its amplification.

\item In parallel, after amplification, the signal beam is sent to probe the spacetime region where the target is located.

\item Classical digital filtering techniques are used in the detection and storage of the received beam.

\item The detection procedure relies on the measurement of linear quadratures and matched filtering to compare both signals. The procedure allows to implement a phase-conjugate receiver that fully exploits the correlations of the two signals without analogue photoreduction (no Matching filtering) technique.
\end{enumerate}

For the classically correlated light illumination, the procedure is similar, with the same conditions of temperature, energy and power for the idler/signal generation than in the case of the quantum illumination. Several artificial sources of noise are added to simulate the noise introduced in the amplification process of the quantum illumination radar (introduced in the quantum radar protocol during amplification of the signal). However, the experiment did not show outperform of the classical benchmark, namely, coherent light illumination in the same conditions of frequency and intensity.

 In the experiment, the target is located up to  $1\,m$ fixed distance from the sending antenna at room temperature. The reflected signal is detected using also heterodyne detection. Then the two measurements are post-processed and used to calculate the signal to noise ratio.

 The JPA generation and amplification was achieved in cryogenic conditions. Indeed,  before amplification, the generation of the microwave modes by the JPA is in a cryogenic container at $7 \, mK$. The two beams generated are at frequencies $\omega_1=\, 10.09\, GHz$ and $\omega_2 =\,6.8\, GHz$), the experiments are done at room temperature, in the sense that the signal is sent to detect the target which is at room temperature.
 For the quantum illumination, $M=380000$, while for coherent light experiments $M=192000$.
\\
 {\bf Experimental results}. The experiment showed an enhancement in SNR when using the quantum entangled microwave radar respect to symmetric, suboptimal classical illumination scheme in the following way:
 \begin{itemize}
 \item Enhancement up to $3\, dB$ in the SNR in the low intensity regime  respect to sub-optimal, symmetric classical illumination.

 \item Under the assumption of perfect idler photo number detection, the SNR gain of quantum illumination is up to $4\,dB$, defeating also coherent light illumination with heterodyne detection (that does not requires phase-conjugate information, over the region where the outputs of the JPA are still entangled.

 \item  For signals with photon number $N_S >4.5$, there is no advantage from entangled to non-entangled classical light and coherent light illumination.
 \end{itemize}
In the region of $N_S<\,0.4$, the experiment results inferred that the gain in SNR in the experiment of quantum illumination respect to homodyne detection using coherent light was $1\,dB$ using phase conjugate receiver respect to coherent illumination using homodyne detection. This is less that the theoretical advantage ($3\,dB$), but is in concordance with theory when taking into account several non-idealities \cite{Barzanjeh et al.2019}.

All these results were applied in the case of symmetric idler/signal intensities.
\subsection{JPA modulation and slow moving target tracking}
 In the above experiments, the target is located at a fixed, known distance. However, the JPA allows for a modulation of the signal and idler frequencies within a narrow range, that could be used to provide a range variable. However, to use this technique, the location of the target has to be known approximately first. For instance, if a slow moving target is detected at a given position using an auxiliary conventional radar, then the frequency modulation of the JPA of a quantum radar prototype can be used to track the position of target in later times \cite{Karsa et al.}. Combined with standard Doppler techniques, it can be used to measure the relative speed of the target too.

 The potentiality of modulation to track slow moving targets could be extrapolated to other forms of entangled generation. Unfortunately, conventional SPDC or SFWM generation are unable to provide modulation in frequency. But such a limitation can be overcome, as for instance discussed in \cite{Walton et al.}.

\subsection{Comparison of quantum noise radar and coherent light illumination}

In reference \cite{Shapiro2019} Shapiro has analyzed the theoretical performance of the protocol discussed in Luong et al. and Barzanjeh et al. and compared with coherent light illumination. Shapiro adopted a correlated source where the intensity can be very unbalanced between signal and idler. This source was previously proposed by Spedalieri et al. in the field of quantum metrology \cite{Spedalieri et al.}, where they showed that a correlated thermal source with a highly energetic idler mode is able to reach the standard quantum limit in the estimation of bosonic losses (or even to surpass this limit in certain non-Markovian assumptions for the environment). In similar way, in  quantum illumination the beams need to have the same intensity (because the way they are generated as composed by entangled beams), but in coherent illumination the relative intensities of the idler signal beams can be arbitrarily distributed. As a result of this variable relative intensity, one can have a weak signal beam classically correlated with a strong idler beam.

Let us consider the hybrid quantum radar protocol as the protocols and prototypes discussed in  \cite{Luong et al. 2019} and in \cite{Barzanjeh et al.2019}. Such hybrid radar systems are named by Shapiro {\it quantum correlated radar}, while the  equivalents protocols at the classical level were called {\it classical correlated radar}. The corresponding correlation matrices show an interplay between the intensities of the idler/signal beams that implies the possibility of outperform the sensitivity conditions achieved by a quantum noise radar by the classical noise radar. Under asymmetric idler/signal system, it is possible to construct coherent signals of the same characteristic than the signal beam of the quantum radar counterpart, while keep a stronger signal idler, due to the remain effect of the {\it classical correlations} in the coherent case when the intensity of the idler beam is large enough  \cite{Spedalieri et al., Shapiro2019}.

 The argument is based on the analysis of the correlation matrices for QCR and CCR in \cite{Luong et al. 2019}. The $M$ modes of the measurement (heterodyne measure) are  a set of independent, identically distributed, complex valued random columns vectors, whose quadrature components have zero mean Gaussian distributions with covariant matrices as follow. For the correlated noise radar in the case of stationary signals, the correlation matrices are $4\times 4$ symmetric matrices that do not depend on time parameter and such that act on four vectors, formally as
  \begin{align*}
  x^\top \equiv (a_{R m},\,\sqrt{N_S+\,1}/\sqrt{G_A}\,a^* _{Im}),
  \end{align*}
  where $G_A$ is the pre-amplifiers gain coefficient.
The correlation matrices can be re-written in the form (compare with the form discussed in \cite{Luong et al. 2019}),
\begin{align}
E^{QCN}_{\eta,\theta}[x\,x^\top]=\,
\begin{bmatrix}
N_R+N_F &  0 & \sqrt{\eta N_S}\cos(\theta) & -\sqrt{\eta N_S}\sin(\theta) \\
0 & N_R+N_F  & \sqrt{\eta N_S}\sin(\theta) & \sqrt{\eta N_S} \cos\theta\\
\sqrt{\eta N_S}\cos(\theta) & \sqrt{\eta N_S}\sin(\theta) & 1+\frac{N_F\,-1}{N_S+1} & 0\\
-\sqrt{\eta N_S}\sin(\theta) & \sqrt{\eta N_S}\cos(\theta)  & 0 & 1+\frac{N_F\,-1}{N_S+1}
\end{bmatrix}
\end{align}
for the quantum correlated noise radar, while for the classical correlated radar, the correlation matrix is of the form
\begin{align}
E^{CCN}_{\eta,\theta}[x\,x^\top]=\,
\begin{bmatrix}
N_R+N_F &  0 & \sqrt{\eta N_S}\cos(\theta) & -\sqrt{\eta N_S}\sin(\theta) \\
0 & N_R+N_F  & \sqrt{\eta N_S}\sin(\theta) & \sqrt{\eta N_S} \cos\theta\\
\sqrt{\eta N_S}\cos(\theta) & \sqrt{\eta N_S}\sin(\theta) & 1+\frac{N_F}{N_I} & 0\\
-\sqrt{\eta N_S}\sin(\theta) & \sqrt{\eta N_S}\cos(\theta)  & 0 & 1+\frac{N_F}{N_I}
\end{bmatrix}
\end{align}
 For the null hypothesis, the correlation matrices are determined by the condition $\eta=0$ (absence of target). The phase $\theta$ is on the interval $0\leq \theta\,\leq 2\pi$. $N_S$ is the average number of photons per mode in the signal beam; $N_I$ is defined analogously for the idler beam. $N_F$ is the noise figure. It is such that $N_F\geq 1$ and it is equal to $1$ in the ideal case. $N_R$ is given by the expression $N_R=\,\eta \,N_S +\,N_B$.

As we remarked before, $N_I$ and $N_S$ are independent variables. Also, it is $N_F$ too. Because of this independence, there are several regimes where classical light outperforms quantum light:
\begin{itemize}
\item For an ideal detector with $N_F=1$, in the limit $N_I\to \infty$, the covariant matrices are identical and hence, the performance of the classical correlated noise radar and quantum correlated noise radar is identical.

\item If $N_F>1$, the classical correlation radar outperforms the quantum analogue when the condition $1+\frac{N_F\,-1}{N_S+1} >\,1+\frac{N_F}{N_I}$ meets. This happens when the idler beam is of such intensity such that
    \begin{align}
    N_I>\,N_F\, \frac{N_S +1}{N_F-1}.
    \label{strong idler condition}
    \end{align}
Note that this result does not depend upon the reflectivity $\eta$.
\end{itemize}
\subsection{Discussion of the scope of Shapiro's  analysis}
The direct consequence from Shapiro's analysis is that, under the conditions investigated, the scheme of hybrid quantum radar with heterodyne detection cannot universally outperform a coherent light radar working with a signal of the same intensity and energy characteristics and under the same detection capabilities. The argument is to retain a bright enough idler, while the signal has the required properties \cite{Spedalieri et al.}. Furthermore, the argument assumes, in the case of microwave beams, that one can generate coherent states at ambient temperatures, which is not clear to be possible.

 This claim partially contrasts  with the results of the experiments performed by \cite{Barzanjeh et al.2019}  demonstrating a modest improvement using entangled light respect to coherent light. They investigated both heterodyne and homodyne detection, but they  used a low intensity idler. Hence, Shapiro's analysis is not in contradiction with the outcome of such experiments.  In addition, note that Shapiro's analysis applies to heterodyne detection, while  in \cite{Barzanjeh et al.2019} reported also experiment with homodyne detection, with a relatively modest increase in sensitive of $1\, dB$ in the signal to noise ratio when using quantum entangled light respect to the strongest classical benchmark as is using coherent light as signal and homodyne detection.

Despite the correctness of Shapiro's argument, for specific tasks, hybrid quantum illumination working with heterodyne detection  could bring technical advantage over coherent light protocols.  The hybrid protocols are of practical interest when the use of coherent light for illumination is impractical. This could be the situation in security applications and in non-invasive scanning applications. On the other hand, it could happen that the power of the idler is not possible to satisfy the condition \eqref{strong idler condition}. Note that the same argument applies generically to arguments involving the comparison of quantum illumination with coherent light illumination, starting with the analysis of Lloyd and Shapiro in \cite{ShapiroLloyd}.

\section{The quantum radar protocol of Maccone and Ren}\label{Maccone Ren protocol}
From the preceding discussion, it is clear that one of the main problems in the application of quantum illumination to radar is what we have been calling  the {\it range target problem}. In particular,  the protocols and prototypes described in \cite{Barzanjeh et al.2019} and in \cite{Luong et al. 2019}, but also in the case of Gaussian quantum illumination \cite{Lopaeva et al.,Zhang et al.,England Balaji Sussman 20019} require exact or approximate knowledge of the target location. Recently, Maccone and Ren have introduced a different scheme to apply quantum entangled light to radar purposes  in a way that corresponds to a type 4 quantum sensor as discussed in section \ref{quantum entanglement and quantum radars} and that does not requires knowledge beforehand of the target range \cite{MacconeRen2019}. We describe briefly below the Maccone-Ren theoretical protocol.
\subsection{Maccone-Ren protocol for quantum radar}
In Maccone-Ren's protocol, quantum states with $N$ entangled photons are prepared. For each individual state, all the $N$ photons are sent to explore a region of spacetime possibly containing a non-cooperative point-like target. The difference with quantum illumination protocols and hybrid protocols \cite{Lloyd2008,Shapiro2019,Tan,Barzanjeh et al.,Barzanjeh et al.2019,Luong et al. 2019,Luong et al. 2019} is that all the entangled photons are sent to explore the target and none is preserved as idler.

In order to introduce systematically these ideas, we use an analogous analysis  as in Lloyd's quantum illumination scheme.
\\
 {\bf A. Protocol for a quantum radar using entangled light}. The entangled state in Maccone-Ren's protocol is an EPR-like state of the form
\begin{align}
|\psi_N\rangle \equiv\,\int\,d\omega\,d\vec{k}\,\psi(\omega,\vec{k})\left( a^\dag (\omega,\vec{k})\right)^N|0\rangle,
\label{entangled states Maccone Ren}
\end{align}
where $ a^\dag (\omega,\vec{k})$ is the creation operator of a photon with frequency $\omega$ and transverse moment $\vec{k}=\,(k_x,k_y)$; the propagation of each of these photons is along the $z$-direction and they have the same momenta $\hbar \,k$. The function $\psi(\omega,\vec{k})$ is the $N$-{\it photon structure function}. For $N=2$, $\psi(\omega,\vec{k})$ is the  {\it biphoton structure function} of this particular class of states.

In what will follow, one assumes  that the far-field approximation
\begin{align*}
 |\vec{k}_3|^2 =\,(k^2_x+\,k^2_y+k^2_z)^2\gg (k^2_x+\,k^2_y )^2=|\vec{k}|^2
 \end{align*}
  holds good.
After scattered by the point object, the joint probability to detect the $N$ photons at times and transverse locations $\{(t_j,\vec{r}_j),\,j=1,...,N\}$ is given by an expression of the form
\begin{align}
p(\{(t_j,\vec{r}_j)_{j=1,...,N}\})\propto\,\left|\langle 0|\prod_j\,E^+(t_j,\vec{r}_j)|\psi_N\rangle\right|^2,
\label{joint probability N photons}
\end{align}
where $E^+(t,\vec{r})$ is the electric field operator in the Heisenberg picture of dynamics at the transverse location $\vec{r}$ and instant $t$ where the photon is detected at the receiver and the scattering is due to a point object located at a transverse distance $\vec{r}_p$.

In the far field approximation, $E^+(t,\vec{r})$ is of the form
\begin{align*}
E^+(t,\vec{r})\propto\,\int\,d\omega\,d\vec{k}\,\int d\vec{r}_0\,\delta(\vec{r}_0-\,\vec{r})\,a(\omega,\vec{k})\,e^{\imath (t-t_0)}\,e^{\imath \vec{k}\cdot (\vec{r}_0-\,\vec{r})},
\end{align*}
where $a(\omega,\vec{k})$ is the annihilation operator of the mode. Then  joint probability of measuring $N$ photons at times and transverse locations $\{(t_j,\vec{r}_j),\,j=1,...,N\}$ is then given by the expression
\begin{align}
p_N(\{(t_j,\vec{r}_j)_{j=1,...,N}\})\propto \,\left|\widetilde{\psi}\left(\sum^N_{j =1}t_j-\, Nt_0,\sum^N_{j=1} \vec{r}_j-\,N\vec{r}_p\right)\right|^2,
\label{probability Maccone-Ren}
\end{align}
where
\begin{align*}
\widetilde{\psi}(t,\vec{r})=\,\int\,d\omega\,d\vec{k}\, \psi(\omega,\vec{k})\,e^{\imath(\omega t+\,\vec{k}\vec{r})}.
\end{align*}
Therefore, the expression for the joint probability of detecting the $N$ photons at times $t_j$ and transverse locations $\vec{r}_j$ can be re-written in the form
\begin{align}
p(\{t_j,\vec{r}_j)_{j=1,...,N}\})\propto \,\left|\widetilde{\psi}\left(N\left(\frac{\sum^N_{j=1} t_j}{N}- t_0\right),N\left(\frac{\sum^N_{j=1} \vec{r}_j}{N}-\,\vec{r}_p\right)\right)\right|^2.
\label{probability Maccone-Ren2}
\end{align}
This expression has several relevant consequences. First, it provides a method to determine the target range.
From the detection times one can extract the target range, which is given by the expression
\begin{align}
r_z=\,c\,\frac{\sum^N_{j=1}(t_j-t_0)}{2\,N},
\label{target range in Maccone-Ren}
\end{align}
where $c$ is the speed of light.
The transverse location of the target is similarly estimated to be given by the average transverse displacement  relation,
\begin{align*}
\vec{r}=\,\frac{\sum^N_{j=1}\,\vec{r}_j}{N}.
\end{align*}
These methodology is a generalization of the coincidence method of radar distance when using non-entangled light $(N=1)$, for instance used by England et al. in their experiments \cite{England Balaji Sussman 20019}.
\\
{\bf B. Protocol for the radar that uses non-entangled light}.
For the experiment where one individual, non-entangled photons are sent to explore and detect the target, the probability of detecting a single photon at time $t$ and transverse position $\vec{r}$ is given by the expression of the form
\begin{align}
p(t,\vec{r})\propto \left|\widetilde{\psi}(t,\vec{r})\right|^2.
\label{probability of one individual photons}
\end{align}
The time of detection transverse location are established by the expectation value of the corresponding coordinates.

\subsection{Enhancement of sensitivity using quantum entanglement respect to non-entangled light}
Since the variance operation of a sum is the sum of the variance, there is a direct comparison between the statistical properties of \eqref{probability Maccone-Ren2} and \eqref{probability of one individual photons}. In particular, there is a suppression factor of order $1/\sqrt{N}$ in the variance of \eqref{probability Maccone-Ren2} respect to the variance of \eqref{probability of one individual photons}.

The reduction in the variance has the following effect. If each individual photon implies an error in the precision of measurement of observables proportional to the standard deviation of \eqref{probability of one individual photons}, then when using instead the states \eqref{entangled states Maccone Ren}, with joint probability of detection \eqref{probability Maccone-Ren2}, the error in the estimation of
the  expected arrival time $t\equiv\,\frac{\sum_j t_j}{N}$ and transversal displacement detection vector $\vec{r}\equiv \frac{\sum_j \vec{r}_j}{N}$ shows a reduction of order $\sqrt{N}$.

Since the range of the object can be defined by the expression \eqref{target range in Maccone-Ren}, one can say that entanglement imprints a reduction in the range detection error of order $\sqrt{N}$. Similar conclusion is established for the expected transverse displacement detection position $\vec{r}\equiv \frac{\sum_j \vec{r}_j}{N}$. When repeated many times the experiment, this advantage becomes enhanced by statistical independence of each individual detected shot.
\subsection{Practical issues implementing Maccone-Ren's quantum radar protocol}
There are several relevant difficulties and issues for  the practical implementation of Maccone-Ren protocol for quantum radar. The first is the inherent difficulty in the generation of the entangled states required \eqref{entangled states Maccone Ren}. A possible partial solution to this important problem is to use partially entangled states, as discussed below. However, note that the benefits of the protocol appear for large $N$, being more and more difficult to produce as $N$ increases.

Second, the randomness in the distribution and arrival in time for the $N$ entangled photons implies an infinite time detection and infinite size of the detector. These two issues are considered in \cite{MacconeRen2019} and cured by using partially entangled states. It is shown that, although the enhancement in precision for the measurement of the range and transverse displacement is not as high as when using the maximally entangled states  \eqref{entangled states Maccone Ren}, the use of partially entangled states of the form
\begin{align}
|\phi\rangle :=\,\int\,d\omega\,d\vec{k}\,\prod_j \,d\omega_j\,d\vec{k}_j \,\psi(\omega,\vec{k})\,\gamma(\omega_j)\,\xi(\vec{k}_j)\,a^\dag (\omega +\omega_j,\vec{k}+\vec{k}_j)|0\rangle .
\label{partially entangled states}
\end{align}
allows for a finite time of detection, finite size of the detection screen. Furthermore, the states \eqref{partially entangled states} are easier to produce than the maximally entangled states \eqref{entangled states Maccone Ren} using SPDC methods, at least for states associated with two entangled photons \cite{Liu et al. 2009, Pires Exter 2009, Yun et al. 2012}. Even if the gain is not as large as using entangled states of the form \eqref{entangled states Maccone Ren}, the use of partially entangled states for quantum radar still has an advantage over classical photon state protocols \cite{MacconeRen2019}.

The third issue in Maccone-Ren's quantum radar protocol is related with the effect of thermal noise, since the entangled states used in the protocol are very sensitive to noise. Indeed, usually protocols in quantum metrology are very sensitive to noise. In the case of Maccone-Ren quantum radar protocol, the lost of one of the $N$ entangled photons renders the other $N-1$ useless, since their detection is produced at random times and transverse locations. It has been suggested several strategies to solve this problem. Two of such suggestions are discussed in \cite{MacconeRen2019}. The first is to use the partially entangled states \eqref{partially entangled states}. Such a states are more robust against noise. The second is suggested by the protocols discussed in \cite{GiovannettiLloydMaccone2002} and involve nested systems of entangled states. Both strategies reduce the effect of noise at the price of a reduction on the enhancement in precision by using entangles states for ranging detection.

\section{Quantum illumination with multiple entangled photons}\label{quantum illumination with multiple entangled photons}
We review in this section a novel  protocol for a quantum radar which is resilient to thermal noise. The protocol scheme is a combination of Lloyd's quantum illumination protocol and Maccone-Ren's quantum radar protocol. Further details are developed in \cite{Ricardo2020b}.

 \subsection{Lloyd's quantum illumination using multiple entangled signals beams}
 We consider here the case where the back-ground noise $N_B$ is small and the time-bandwidth product $M$ is large, as in the original analysis from Lloyd. This has the advantage of simplifying the treatment.

 The generation of the idler and signal beams are as follows.
We first consider a state of the form
\begin{align}
|\psi\rangle_3 =\,\frac{1}{\sqrt{M}}\,\sum^{M}_{\alpha=1}\,\hat{a}^\dag(\omega_1(\alpha),\vec{k}_1(\alpha))\,\hat{a}^\dag(\omega_2(\alpha),\vec{k}_2(\alpha))\,\hat{a}^\dag(\omega_3(\alpha),\vec{k}_3(\alpha))\,|0\rangle,
\label{three generation}
\end{align}
where $\alpha =1,...,M$ indicates the modes of the states.
This photon state can be obtained by four interaction linear optics.
The photons $1$, $2$, $3$ are correlated in time (they are generated at the same time) and they are correlated in frequencies by the relation $\omega_0=\,\omega_1+\omega_2+\omega_3$ and in momentum, by the relation
$\hbar\vec{k}_0=\,\hbar\vec{k_1}\,+\hbar\vec{k}_2\,+\hbar\vec{k}_3 $  (see for instance, \cite{GarrisonChiao}, section 13.4), where $(\omega_o,\hbar\vec{k}_0)$ are the frequency and momentum of the pump beam.
The initial state can be transformed by an unitary operator $\mathcal{U}=\,I_1\otimes I_2\otimes U_3$ that sends $\frac{\vec{k}_3}{\|\vec{k}_3\|}$ to $ \vec{k}'=\frac{\vec{k}_2}{\|\vec{k}_2\|})$ to prepare the sates in the form
\begin{align}
\widetilde{|\psi\rangle}_3 =\,\frac{1}{\sqrt{M}}\,\sum^{M}_{\alpha=1}\,\hat{a}^\dag(\omega_1(\alpha),\vec{k}_1(\alpha))\,\hat{a}^\dag(\omega_2(\alpha),\vec{k}_2(\alpha))\,\hat{a}^\dag(\omega_3(\alpha),\vec{k}'_2(\alpha))\,|0\rangle,
\label{three generation collimated}
\end{align}
where two of the photons have parallel momenta $\vec{k}_2 \, \| \,\vec{k}'_2$, but were the frequencies $\omega_2$ and $\omega_3$ can be different. After this preparation,
the beam is split into two beams: the idler beam, which is composed by photons with momentum $(\omega_1,\vec{k}_1)$, and the signal beam, which is composed by states with two photons with momentum $(\omega_3,\,\|\vec{k}_3\|\,\vec{k}'_2)$.
The signal and the idler state are not any more entangled, but mixed states. However, the following correlations persists:
\begin{itemize}
\item The time correlation in the generation of the three photons,

\item The photons on each state exhibit correlations in energy,
\begin{align}
\hbar\omega_0=\,\hbar\omega_1(\alpha)+\,\hbar\omega_2(\alpha)+\,\hbar\omega_3(\alpha),\quad \alpha =1,...,M
\label{correlation relation 1}
\end{align}
\end{itemize}
The signal state is denoted by $\tilde{\rho}_2$, while the idler state is denoted by $\tilde{\rho}_1$. They can be both mixed states.
The noise is described by the state $\rho_0$ given by the relation \eqref{state for noise} as in Lloyd's theory. Note that although the photons are initially correlated in momenta, such correlation is lost during the preparation of the signal and idler beams.

The criteria that we can follow for a positive detection is formulated as follows:
\bigskip
\\
{\bf Criterion for positive detection:}{\it  We declare that the target is present if two photons in the spectrum range of the signal are detected back within an established time window at the same time than an idler photon is detected together in a joint measurement of the idler and signal beam and if the correlation relation \eqref{correlation relation 1} holds good.}
\bigskip
\\
The precise notion of time window detection must be adapted to the particular experimental setting. It cannot be too large to confuse uncorrelated photons, but also not to short to miss entangled photons.

The evaluation of the signal to noise ratio for non-entangled illumination and for quantum illumination with multiple entangled photon states as signal has been carried out in \cite{Ricardo2020b}, following an analogous procedure as in \cite{Lloyd2008}.
\\
{\bf A. Illumination with non-entangled light}. When the target is not  there and the illumination is done with non-entangled light, the quantum state is described by a density matrix of the form
\begin{align*}
\rho_0  \approx \left\{(1-\,M\,N_B)|0\rangle\langle 0|+\,N_B\,\sum^M_{k=1}\,| a^\dag (\omega_1,\vec{k}_1)|0\rangle\,\langle 0|a(\omega_1,\vec{k}_1)|\right\}.
\end{align*}
This is the noise state used in Lloyd's theory (see the expression \eqref{state for noise} in Appendix \ref{AppendixLloyd quantum illumination}). The probability of false positive  can be read directly from the structure of the state and, by the criteria of detection discussed above, it is the probability of detecting two photons in the same time window. Therefore, in the case of low bright environment ($N_B<<1$), the probability of false positive is given by the expression
\begin{align}
p_0(+)=\,(N_B)^2.
\label{N=3 non-entangled light nothere}
\end{align}
On the other hand, when the target is there, and under the same assumptions, the state is given by the density matrix
\begin{align*}
\rho_1& =\,(1-\eta)\rho_0+\eta\tilde{\rho}\\
& \approx (1-\eta)\left\{(1-\,M N_B)|0\rangle\langle 0|+\,N_B\,\sum^M_{k=1}\,| a^\dag (\omega,\vec{k})|0\rangle\,\langle 0|a(\omega,\vec{k})|\right\}+\,\eta \,\tilde{\rho} ,
\end{align*}
where $\tilde{\rho}$ stands for the state describing the signal when using non-entangled light.
The probability of simultaneous detection, according to our criteria of target there if two photons are detected, is in this case of the form
\begin{align}
p_1(+)=\,((1-\eta)N_B+\,\eta)^2
\end{align}
The signal to noise ratio in quantum illumination with two signal state photons when the illumination is performed with non-entangled light is given by the expression
\begin{align}
SNR_{CIMR}=\,\frac{p_1(+)}{p_0(+)}=\,\frac{((1-\eta)N_B+\,\eta)^2}{(N_B)^2}.
\label{SNR noentanglement in Maccone-Ren illumination}
\end{align}
One observes that this signal to noise ratio is given by the square of the signal to noise ratio $SNR_{QI}$ in Lloyd's theory (expression \eqref{SNR noentanglement} in Appendix \ref{AppendixLloyd quantum illumination}). Therefore, using the above criteria for detection and using classical illumination, reduces considerably the SNR respect to the usual criteria of positive only if a photon is detected.
\\
{\bf B. Illumination with four interaction generated entangled light}.
When there is no target present, the state noise-idler is described by the density matrix $\tilde{\rho}^e_0$ and has the form
\begin{align*}
\tilde{\rho}^e_0 \approx \,\left\{(1-\,M\,N_B)|0\rangle\langle 0|+\,N_B\,\sum^M_{k=1}\,| a^\dag (\omega,\vec{k})|0\rangle\,\langle 0|a(\omega,\vec{k})|\right\}\otimes\,\tilde{\rho}_1.
\end{align*}
The probability of a false positive is the probability to attribute to the presence of the target the detection of two simultaneous returned photons. Within the scope of the approximations that we are considering, such a probability is independent of the details of the signal state and given by the expression
\begin{align}
p^e_0(+)=\,\left(\frac{N_B}{M}\right)^2.
\label{probability of error multiple photon}
\end{align}
This relation shows an enhancement respect to the analogous relation in Lloyd's quantum illumination protocol (eq. \eqref{entanglemet0+} in Appendix \ref{Lloyd quantum illumination}).

When the target is there, the state after decoherence and interaction signal-target is of the form
\begin{align*}
\tilde{\rho}^e_1 =\,(1-\eta)\cdot \tilde{\rho}^e_0+\,\eta\,\tilde{\rho}_2,
\end{align*}
By a similar argument as in Lloyd's theory, the probability of detection using entangled light signal states when the target is there for one trial is
\begin{align}
p^e_1(+)=\,\left((1-\eta)\,\frac{N_B}{M}+\eta\right)^2.
\end{align}
The signal to noise ratio is of the form
\begin{align}
SNR^e_{QIMR}=\,\frac{p^e_1(+)}{p^e_0(+)}=\,\left(\frac{M}{N_B}\,\right)^2\,\left((1-\eta)\frac{N_B}{M}+\eta\right)^2,
\label{SNR entanglement Maccone Ren}
\end{align}
which is the square of the signal to noise ratio obtained for Lloyd's quantum illumination in the analogous case, equation \eqref{SNR entanglement}.

Expression \eqref{SNR entanglement Maccone Ren} reflects the enhancement in sensitivity of the use of quantum entangled states respect to non-entangled states. Furthermore, the probability of error \eqref{probability of error multiple photon} is reduced respect to Lloyd's quantum illumination, but  the signal to noise ratio is reduced respect to Lloyd's quantum illumination.
\subsection{Reduction of the required time band-width product}
The first advantage of the multiple entangled photon quantum illumination protocol is a reduction
the time-bandwidth product required to have enhancement can be reduced respect to usual quantum illumination. This property can have implications as a potential remedy for the problem discussed in discussed in Section \ref{problems of the microwave qi} in the context of the applicability of microwave quantum illumination. An account of this property can be find in \cite{Ricardo2020b}.
\subsection{Determination of the range and transverse position using quantum illumination  with two signal photon states} The second advantage of the protocol over standard quantum illumination protocols is that it provides a method for the determination of the target range.
If the target is small enough, the criterion for the detection of a stealth target is that, as discussed before, the target is declared detected if two individual photons with the same frequency and momenta are detected within the same detection time window. Under the further assumption that there is only a pair of photons on fly, the detection of a pair of correlated photons provides a measure also of $t-t_0$ and determines the range by the expression
\begin{align}
r_z=\, c\,\frac{1}{2}\left(T-t_0\right)
\end{align}
where $T$ is the arrival time (it is assumed that both photons arrive simultaneously, modulo the time window error) and $t_0$ is the time of emission.
The measurement of the location of the two photons determines the transverse location of the target as the average location of the photons arrivals.

The above strategy to determine the range could also be applied to Lloyd's quantum illumination protocol (this is the methodology used for instance in \cite{England Balaji Sussman 20019}), but the precision of the method is smaller than when using signal beams with two photons. Note that the net effect of the two photon signal states is to mark with an extra-reinforcement in the correlations, the signals coming from the scattering with the target: two photon signal states are necessary to determine the transverse position.

Finally, let us note that other methods for reception can be used, as for instance hybrid methods, that do not require to keep the idler alive.
\subsection{Potential issues with the implementation of the method}

In order to use quantum illumination with multiple entangled photons in long range radar applications, the signal beam must be generated in the microwave regime. However, the original entangled state $|\psi_3\rangle$ does not need to correspond to a state in the microwave regime. In general will be a superposition of wave packets with center in the optical regime. Existing current frequency conversion methods \cite{Barzanjeh et al.} can be applied to down convert frequencies. The application of such methods will reduce the efficiency drastically. On the other hand, advances in quantum sensing in the terahertz spectrum \cite{Kutas et al.2019} are potentially extended to the microwave regime.

Another technical problem in the implementation of the protocol proposed in this section is that the idler photon needs in principle to be stored and keep alive the idler beam, in order to perform a quantum measurement when the two pairs of beams arrive, since it is fundamental to keep track from the time correlation. Either advances in idler storage or digital techniques as the ones used in \cite{Barzanjeh et al.2019,Luong et al. 2019} are required.

\section{Other proposals for quantum radar}\label{Other protocols}
The protocols and prototypes  for quantum radar already discussed do no exhaust all the proposals for the use of quantum entanglement for radar purposes discussed in the last years. In this section we will discuss in some detail another proposal for quantum radar that address the range problem. This is the proposal from Durak et al. \cite{DurakJamDindar2019}. We have considered the proposal from Durak et al. because it is very close to current protocols for standard radar systems and hence, it can be replicated and extended in a easier way.

Other proposals that have recently appeared in the literature and that also aim solutions for the range problem are discussed in  \cite{Frick McMillan Rarity 2020}, where they explode the properties of mixed squeezed states to enhance sensitivity. Another proposal is  discussed in \cite{He et al. 2020}, although the second of them only address the presence/non-presence of the target. Although these and other investigations are very interesting as attempts to solve the range problem, we will not consider here in detail.
\subsection{Quantum radar proposal of Durak, Jam and Dindar} In the work \cite{DurakJamDindar2019} by Durak, Jam and Dindar  a protocol for quantum radar has been discussed which is in principle capable to provide the range of the target without the need of previous knowledge of it. The protocol is a realization of Lloyd's quantum illumination protocol, where the signal and the idler photons are correlated in frequency, polarization and detection time. Each signal photon of the entangled pair is sent to explore the region of interest. The scattered signal is received first through a telescope, then an avalanche single photon detector is used to detect the photon and the event is registered. The distribution of the received time detections is described by the function $D_R(t+\tau)$. The photons in the idler beam are detected using single avalanche detector and time recorded just after generation with a time distribution of the form $D_I(t)$.  The protocol assumes than only photons entangled in polarization with the photons of the signal beam will arrive, which is a valid hypothesis for the frequencies considered.

By using cross-correlation model between the timing of arrival and the Doppler frequency shift, Durak et al.  discussed a methodology to obtain the location of the target and its speed. The cross correlation is of the form
\begin{align*}
ccf (\tau, \gamma, f_d)=\,K\,\int^T_0\,D_I(t)\,D_R(t+\tau,\gamma,f_d)\,dt,
\end{align*}
where $\gamma=\,\frac{1}{\sqrt{1-\beta^2}}$ is the relativistic $\gamma$-factor, and $f_d$ is the Doppler shift of the received signal. The detection of the target is determined by the time of the pick of the cross-correlation, while the velocity is found by measuring the Doppler frequency shift at the time $f_d$ of the correlations.

One of the theoretical benefits of Durak et al. protocol is that in principle, it can be used at any signal frequency, provided the detection requirements are available and several of the hypothesis done hold, for instance, the hypothesis on the conservation of polarization. If the Durak et al. can indeed be used at microwave frequencies, it could potentially be significant for radar applications.
\subsection{Proof of principle experiment}
Durak et al. reported an experimental demonstration of the protocol through an experiment with the following characteristics. The signal/idler entangled photons are obtained by parametric down conversion where the pump beam is at $\lambda_0 =\,402\, nm$ and the down conversion is to wavelengths $\lambda_1=\,780 \,nm$ and $\lambda_2=\,842\, nm$.  After passing through a filtering and collimation processes, the idler and the signal are also correlated in polarization. Then the signal is sent through a telescope of aperture $50\, mm$ to explore the location (in the experiment the range of the target is fixed) of a target composed by black anodized Aluminum. The scattered photons are detected back by the telescope, filtered and the arrival time registered using single photon counting detector. The photon detector used allowed to count $5 \times \,10^5$ pairs of photons per second.

In the reported experiment, the target is situated at a fixed distance from the telescope. The time delay in the arrival of the signal photon respect to the idler is theoretically modeled by a curve of the form $f(r)=b+a\,R^{-2}$, where $R$ is the target range. This model is  based upon the principle that the power detected by the telescope is of the form $P\sim \,D^2/R^2$, where $D$ is the aperture of the telescope. The fit parameters $a$, $b$ depend on the telescope aperture.

 The paper of Durek et al. reports  a good performance of the experimental quantum radar prototype up to a range of $200\, mm$. This maximal experimental range highly depends upon the telescope aperture, the detector time jitter and the power of the source of entangled photons. Improvement in these parameters would imply an extension of the maximal target range.
\subsection{Critical view of  Durak et al. protocol}
Several comments are in order.
First, we would like to remark that in the proof of concept experiment described in  \cite{DurakJamDindar2019} the comparison with the equivalent classical prototype was not discussed; only it was indicated that the experimental results agree with theory. Hence one cannot claim enhancement from such demonstratio, the claim in the paper based upon the agreement between theory and experimental data. Without this direct comparison, it is difficult to establish the enhancement of quantum radar respect to standard radar.

The proposal from Durak et al.  is based in a weak notion of entanglement, similar to the working definition of entanglement used by Chang et al. \cite{Chang et al. 2019}. These notions of quantum correlations may have a sound theoretical foundation in the notion of {\it quantum discord}, briefly discussed in the Glossary, and in the notion of {\it enhancement entangled breaking channels}, according to Sacchi \cite{Sacchi2005a,Sacchi2005b}. Also, in order to implement Durak et al.  protocol for a realistic radar, the signal must work on the microwave regime, where the difficulties of detection of individual photons and generation of the signal are acute. Electro-optomechanical conversion could be a method to be implemented, but at the expenses of reducing the efficiency and introducing further noise.

Increasing the number of detection pairs per second implies that the window for photo detection must be smaller. Currently, there are avalanche photodiode detectors (APD) with a coincidence window (that we can take as the time jitter of the detector) of order $80$ ps. This scale  allows for detection of $\, 10^{10}$ photon pairs per second. Increasing the capability to detect more pairs per second implies that powerful sources can be used for the illumination and hence, the range of the radar can be expanded. Currently, APD with a time jitter of approximately $10\, ps$ are being investigated \cite{Rothman et al. 2019}, which will imply an increase in the performance of the quantum radar protocol. For APD detectors with a time jetter of order $10\, ps$, the range will correspond up to $300 \,m$. However, these figures are still far from the upper bound on the correlation time $t_c \sim \,10^{-1} ps$ in the spontaneous down conversion \cite{Burnham and Weinberg}. Hence improvement in the maximal range can be expected, although eventually will be limited by fundamental physical synchronization of the photon entangled pairs. This fundamental limit is related with a maximal target range of around $30 \, km$ or more (since $t_c$ is an upper bound).

In conclusion, although the method proposed by Durak et al. is interesting from a practical point of view, it is necessary further studies to determine the existence of enhancement respect to classical illumination.

\section{Conclusions and outlook}\label{Conclusion and Outlook}
Quantum radar refers to several protocols and prototypes that, either theoretically or in a preliminary experimental phase, aimed to explode quantum entanglement or quantum correlations properties for enhancement in target detection sensitivity. Although quantum radar is a promising area of research, the realization of a real quantum radar prototype remains elusive. Indeed, the acknowledged large gap between the first theoretical expectations and the subsequent better theoretical understanding and  experimental results is a source for the skepticism on the real possibilities of  enhancement of a quantum radar respect to conventional radar in practical situations. This criticism has been raised especially in the case of quantum illumination, since quantum illumination raised very high expectations on its possible long range radar applications, expectations that gradually have been relaxed. Indeed, it is remarkable that in the scientific literature, there is no reference to a genuine, real quantum radar or range finding  prototype based on quantum illumination.

Despite this initial deception, the benefits that a quantum radar technology can offer over conventional radar technologies still constitute a strong motivation to understand better the applicability of quantum radar protocols. Experiments show \cite{Lopaeva et al.,England Balaji Sussman 20019} that quantum radar implies an enhancement respect to classical illumination and it is only potentially beaten by protocols or prototypes that use  coherent for  illumination. It is also true that these experiments work under assumptions and conditions very different than long range radar. These experiments suggest that there is no an universal quantum radar, but that quantum radar based upon quantum illumination protocols can be useful for some specific tasks: the ones where the use of coherent light is precluded for different reasons.

We have seen that the physical realization of the quantum radar protocols discussed in the literature is precluded by the existence of several issues.
\bigskip
\\
{\bf General problems affecting quantum radars}.
\begin{itemize}
\item The target range problem. As we have discussed, this problem affects to any quantum radar protocol. We think that this problem is among the most pre-eminent problem to prevents the realization of a realistic quantum radar. In principle, the protocols and prototypes discussed by Maccone and Ren \cite{MacconeRen2019}, Durak et al. \cite{DurakJamDindar2019} and the authors \cite{Ricardo2020b} provide different solutions for the range problem.

\item Loss of intensity in the signal beam by attenuation processes. This is a problem that affects  all types of quantum radar protocols. In the context of quantum illumination has been examined recently by Sorelli et al. \cite{Sorelli et al.}. The conclusion reached by their analysis is that, in the microwave regime, the power used in signalling is not enough for target detection, in the regime $N_S<< 1$ of quantum advantage.
\end{itemize}
\bigskip
{\bf Problems for any quantum radar based upon quantum illumination}.
\begin{itemize}

\item Idler storage. In order to make joint measurements or in general, to correlate the idler with the received signal beam, the information provided by the idler must be handle. Two ways have been discussed: 1. The use of very efficient idler storage systems, including quantum memories \cite{Barzanjeh et al.}, 2. The use of classical digital methods and matched filtering \cite{Barzanjeh et al.2019,Luong et al. 2019,DurakJamDindar2019}. Currently, these methods introduce a reduction in maximal target range  and in the SNR enhancement.

\item Detection. For quantum illumination, there is an ideal detector, with a maximal possible gain of $6\, dB $, the FF-SFG detector \cite{Zhuang Zhang Shapiro b}. But current technological limitations and inefficiencies are drastically reduces the prospects to explode the theoretical $6\,dB$ advantage. Direct photo detection is used in several proposals and experiments \cite{Lopaeva et al.,England Balaji Sussman 20019,DurakJamDindar2019,Ricardo2020b}. But the photo counting rates limits the range of the detection significatively \cite{DurakJamDindar2019}.

\end{itemize}
\bigskip
{\bf Problems for the implementation of a quantum radar based on quantum illumination at microwave frequencies}.
\begin{itemize}

    \item Related with this problem is the time-bandwidth problem at microwave frequencies. While at optical frequencies it is possible to reach high products, the situation is much more difficult at microwaves frequencies \cite{Shapiro2019}, the ones used in long range radar.

\item Generation of entangled beams. The usual procedure for generation of quantum entanglement is SPDC. While this is a procedure quite useful for optical frequencies, in the range of microwave frequencies it has several problems. Direct methods of entangled microwave generation like JPA generation require a very low cryogenic temperature \cite{Barzanjeh et al.2019,Chang et al. 2019,Chang et al.2018,Luong et al. 2019}, while frequency conversion  \cite{Barzanjeh et al.} is currently highly inefficient. This is a point that can be greatly improve if new methods to generate quantum entangled states are discussed briefly in Appendix \ref{non-linear optics section}. Other protocols requires the generation of states with  three entangled photons \cite{Ricardo2020b} or states with specific momentum correlation \cite{MacconeRen2019}, which are also difficult to achieve for radar purposes, specially in the microwave regime.
\end{itemize}
\bigskip
{\bf Problems for the implementation of a quantum radar based on quantum illumination at optical frequencies}.
\begin{itemize}
    \item  The target fading problem, that happens when one does not have full knowledge of the target’s reflectivity and phase, is another issue that affects quantum illumination, specially in the optical regime \cite{ZhuangZhangShapiro}.

\end{itemize}

The points discussed above currently prevent the realization of an ultimate, universal quantum radar. However, different protocols can potentially have advantages respect to classical illumination protocols and prototypes in specific applications, specially in noise, entanglement breaking channels, non-invasive applications and in space target detection.

In particular, the target range problem is specially relevant in quantum illumination and quantum interferometric radar \cite{Smith 2009} and needs to be handle specially. On the other hand, when the range is approximately known, quantum illumination can be used in noise resilient, non-invasive scanning systems. For the protocols where the range problem is theoretically partially or totally solved, as in Maccone-Ren protocol \cite{MacconeRen2019}, Durak at el protocol \cite{DurakJamDindar2019} and the protocol discussed in section \ref{quantum illumination with multiple entangled photons} \cite{Ricardo2020b}, the main problems are related with the generation of the specific entangled states, losses and in the idler storage. Possible ways to overcome the detection limitations could be achieved by using classical digital techniques as in the hybrid prototypes  \cite{Luong et al. 2019} and \cite{Barzanjeh et al.2019}.

 Other related protocols that have been recently investigated are \cite{Frick McMillan Rarity 2020}, that consider anti-correlation in frequencies in thermal SPDC production to identify correlated photons, a methodology that is used for range finding. Another protocol has been considered in \cite{He et al. 2020}, where they use recent advances in tunable SPDC non-classical light to enhance sensitivity in detection.

The other general form of quantum radar that has been intensively investigated are protocols where quantum entanglement is preserved during the round trip, specifically in quantum interferometric radar protocols \cite{Marco Lanzagorta 2011,Boto et al.,GilbertHamrick,GilbertWeinstein2008,Smith 2009}. The problems discussed above for entanglement breaking quantum channels protocols also affect quantum interferometric quantum protocols. Losses and noise have even more dramatic effect on quantum interfereometric radar, because the phenomena of quantum decoherence and attenuation. Based on these considerations, quantum decoherence effects will provide a very short maximal target range detection, as it is discussed for instance in the Glossary. This contrast with the claims in \cite{Marco Lanzagorta 2011}, section 5.2 and in \cite{Smith 2009}.

As a conclusion, the above discussions show the many problems to realize a realistic quantum radar. But these difficulties in the practical implementation can be seen as an opportunity to apply and develop quantum technology in the field of range detection.
\newpage

\appendix

\section{Glossary}\label{Glossary}
In this {\it Glossary} we have collected and discussed several theoretical notions of quantum mechanics, quantum optics and quantum metrology that will facilitate the reading of the main part of the manuscript for a non-expert in quantum optics reader. The chosen collection of concepts and results discussed cannot be in any instance comprehensive  and depends on the taste and interest of the authors, but we hope that it will contribute tothe understanding of the text.

Most of the items are compiled here from other sources, except the discussion of the application of the model of decoherence for localization of the wave function, which we have developed an original approach to the problem.
\\
{\bf 1. Pure and mixed states in quantum mechanics.} In the standard formulation of quantum mechanics \cite{Dirac1958}, the state of a physical system $\mathcal{S}$ is completely specified when a normalized element $|\psi\rangle$  of the Hilbert space $\mathcal{H}$ is given. In Dirac bra-ket notation, normalized means that the relation $|\langle \psi|\psi\rangle|=1$ holds good. It is said that the a physical system described by a $|\psi\rangle$ is a pure state. According to the Copenhagen interpretation of quantum mechanics, a pure state provides the most complete description of the system: no further experiments with identically prepared states can provide a more precise description.

    A mixed state of a physical system is a state that is not a pure state. This means that, given the description of the state, one can at least ideally conceive an experiment that will provide a more complete information (hence, description) of the system. Since the complete description of physical systems is in terms of normalized elements of the Hilbert space, the description of a mixed state must be given in terms of them, but cannot be associated to a particular normalized element of $\mathcal{H}$. Indeed, the description of a mixed state is associated to a collection
    $\{(|\psi\rangle,p_i)\}_{i\in I}$, where  $|\psi_i\rangle\in \,\mathcal{H}$ such that
    \begin{align}
    p_i > 0 \,\, \forall\, i\in \,I;\quad \sum_{i\in\,I} p_i =\,1.
    \label{condition probabilities 2}
    \end{align}
The collection $\{(|\psi\rangle,p_i)\}$ is usually interpreted as describing an statistical ensemble, where some elements are described by $|\psi_1\rangle$, some elements by $|\psi_2\rangle$, etc... with corresponding probabilities $p_1$, $p_2$,... Then the expected value of an observable is given by the expression
\begin{align}
\langle X \rangle =\,\sum_i \,p_i \langle \psi_i |\,X\,|\psi_i\rangle =\,Tr [\hat{\rho} \,X],
\label{expentation value}
\end{align}
where the {\it density matrix} is
\begin{align}
\hat{\rho} =\,\sum_i\,p_i\,|\psi_i\rangle \langle \psi_i |.
\label{density matrix}
\end{align}
Every pure state described by a ket $|\psi\rangle$ can also be described by  a density matrix of the form $\hat{\rho}_{|\psi\rangle} =\,|\psi\rangle \langle \psi|$.
In general, a linear operator $\hat\rho:\mathcal{H}\to\,\mathcal{H}$ with the following characteristics (see \cite{Holevo 2011}, section 1.6):
\begin{itemize}

\item It is an Hermitian operator, $\hat\rho =\,\hat\rho^\dag$,

\item It is positive,
\begin{align*}
 \langle \psi |\hat\rho|\psi\rangle\geq 0,\quad \forall\, |\psi\rangle \in\,\mathcal{H},
 \end{align*}

\item $\hat\rho$ it is unit trace, $Tr[\hat\rho]=1$.

\end{itemize}
The statistical properties associated with these operators are formally the same than the statistical properties that one can formulate for a density operator $\hat\rho_{|\psi\rangle}$ associated to a quantum state $|\psi\rangle\in\,\mathcal{H}$ \cite{Holevo 2011}.
Conversely, given a density matrix $\rho$, it corresponds to a pure state iff
\begin{align*}
Tr[\hat{\rho}^2]=1
\end{align*}
holds good.
Indeed, a measure of a mixing of an ensemble system is given by the {\it purity},
\begin{align}
\mathfrak{P}[\hat{\rho}]=\,Tr[\hat{\rho}^2].
\end{align}
For a pure state, the purity is maximal and equal to $\mathfrak{P}[\hat{\rho}_\psi]=1$. The maximally mixed state is found for ensembles where either $p(\psi)=0$ or $\pi(\psi)=a$, a constant independent of $\psi$ in the decomposition of $\hat{\rho}$.

Another measure of the mixing of a state given by a matrix density is given by von Neumann entropy,
\begin{align}
\mathfrak{S}=\,- Tr[\hat{\rho}\ln\,\hat{\rho}].
\end{align}
For a state of the form \eqref{density matrix} subjected to the condition \eqref{condition probabilities 2}, Von Neumann entropy is zero for a pure state and is maximal for a maximally mixed state.

 The basic properties of the density matrix can be found for instance \cite{GarrisonChiao}. General treatments of the theory of density matrix can be found in  \cite{Holevo 2011,von Neumann 1933 version 1955}.
\\
{\bf 2. Schmidt's decomposition}. Given the Hilbert space $\mathcal{H}\cong\,\mathcal{H}_1\otimes \,\mathcal{H}_2$ and a general element $|\psi\rangle\,\in\,\mathcal{H}$ of the form
\begin{align*}
|\psi\rangle =\,\sum_{i,j}\,\psi_{ij}\,|\Phi_i\rangle\otimes |\eta_j\rangle,
\end{align*}
where $\{ |\Phi_i\rangle\otimes |\eta_j\rangle \}$ determines a basis of the product space $\mathcal{H}_1\otimes \mathcal{H}_2$. Then there is an alternative decomposition of $|\psi\rangle$ in terms of product spaces such that
\begin{align*}
|\psi\rangle =\,Y_1\,|\xi_1\rangle\otimes |\vartheta_1\rangle+\,|\psi_1\rangle
\end{align*}
such that the coefficient $|Y_1| $ is maximal. Following this procedure, for finite dimensional Hilbert spaces $\mathcal{H}_1$, $\mathcal{H}_2$, there is a decomposition of $|\psi\rangle$ in product states of the form
\begin{align}
|\psi\rangle =\,\sum^r_{n=1}\,Y_n\,|\xi_n\rangle\otimes |\vartheta_n\rangle,
\label{Schmidt's decompostion}
\end{align}
where $r\leq \,min(\dim(\mathcal{H}_1),\dim(\mathcal{H}_2))$. The minimal value $r=1$ occurs when $|\psi\rangle $ is a product state. Note that the set $\{|\xi_n\rangle\otimes |\vartheta_n\rangle\}^r_{n=1}$ in the Schmidt's decomposition depends upon the initial state $|\psi\rangle$.
\\
{\bf 3. Positive partial transpose criterion}.
The density matrix of a bi-partite separable system is of the form
\begin{align}
\rho=\,\sum_A \,\omega_A\,\rho_{1A} \,\otimes \, \rho_{2A}.
\label{separability of a mixed state}
\end{align}
Then the partial transpose
\begin{align}
\sigma=\,\sum_A \,\omega_A\,\rho_{1A}^\top \,\otimes \, \rho_{2A}
\end{align}
must also describe a separable state.

From the properties of the density matrix, one has that the transpose matrix $\rho_{1A}^\top =\,\rho_{1A}^*$ is also non-negative and has unit trace. Therefore, $\rho_{1A}^\top$  can also be a legitimate density matrix for the subsystem $1$. This implies that all the eigenvalues of the partial transposed matrix $\sigma$ must be non-negative. This algebraic requirement is the {\it positive partial transpose criteria} \cite{Peres1996, HorodeckiHorodeckiHorodecki}. If the criteria is not met for a particular density matrix, then the system is non-separable. This can be the case even if Bell's inequalities are satisfied \cite{Werner1989}. Hence the partial transpose criteria is stronger than Bell's inequalities as a necessary criteria for separability. For two mode gaussian entangled states, a detailed development is known \cite{AdessoIlluminati2007, Weedbrook et al. 2012}.
\\
{\bf 4. Classical and Quantum Chernov's bounds}. The classical result of H. Chernov concerns the problem of given a set of measurements $\{X_i,\,i=1,2,...,N\}$ in the measure of $N$ identically distributed random variables, to determine which is the probability distribution between two possible options $p_1$ and $p_2$ with the minimal possible error. The probability of error $p_e$ is the probability to obtain the hypothesis 0 under the condition that the correct hypothesis is $1$ plus the probability of choosing hypothesis 1 under the condition that the correct hypothesis is 0.

 Chernov solved this problem in the case of asymptotically large trials $N\to +\infty$ \cite{Chernov}, that showed that the probability of error $p_e$ in discriminating among $p_1$ and $p_2$ decrease exponentially with the number of trials $N$,
\begin{align*}
P_e \sim \, \exp (-\,C\,N).
\end{align*}
 The exponent $C$ is known as the Chernov's distance, Chernov's entropy or Chernov's bound. It is determined in Chernov's theory by the probability distributions $p_1$ and $p_2$.

 Although Chernov's theorem is a seminal result in classical decision theory, this type of problems also appear in quantum mechanical problems. Of special relevance is the the discrimination problem between quantum states or discrimination of quantum canals. Therefore, a quantum version of Chernov's theorem and theory appears as a fundamental piece in the theory of quantum discrimination. There are several different generalizations at the quantum level. The following formulation of the quantum Chernov bound makes uses of a {\it positive operator valued measure} (POVM), that consists in this case of two operators $E_0,E_1$ such that $E_0+\,E_1=\,\mathbb{I}$ and $E_i \geq 0$. Given two possible quantum states with corresponding a priori assigned probabilities $\pi_0,\pi_1$, the error probability is given by the expression
 \begin{align}
 p_e = \,\pi_0\,Tr [E_1 \,\hat{\rho}_0]+\,\pi_1 \,Tr[E_0\,\hat{\rho}_1].
  \end{align}
 The minimum of this error probability is denoted by $p_{e,min}$. The basic problem to solve is to understand how the probability of error behaves for $N$ experiments where the state can be either in state $\hat{\rho}_0$ (hypothesis $H_0$) or in the sate $\hat{\rho}_1$ (hypothesis $H_1$). The resolution in the asymptotic case of $N$ determines the quantum Chernov's theorem, which states that \cite{Audenaert et al.}
 \begin{align}
 p_{e,min,N}\sim\, \exp(-N\,C_q),
 \end{align}
 where the exponent is given by the expression
 \begin{align}
 C_q =\,\lim_{N\to \infty}\,-\frac{\log (P_{e,min,N})}{N}.
 \end{align}
This expression of the quantum Chernov bound reduces to the classical result \cite{Audenaert et al.}.

The quantum Chernov's bound is in general difficult to be determined. Other related notions are used in quantum illumination literature, specifically the quantum version of Bhattacharyya bound \cite{Kailath 1967,Pirandola Lloyd} is extensively used the theory of quantum illumination \cite{Tan}. Bhattacharyya bounds approaches to Chernov's bound when the difference in the states $1$ and $2$ is small.

 Also related, is the notion of quantum channel discrimination \cite{Sacchi2005a,Sacchi2005b}, which is on the basis for the initial studies on quantum illumination \cite{Lloyd2008}.
\\
{\bf 5. Quantum decoherence}. Quantum decoherence is the process by which quantum coherence is lost by interaction between the quantum initially coherent system and the environment. Originally motivated by the problem of measurement in quantum mechanics \cite{Zeh1970, Zurek2002}, the theory of quantum coherence has become of fundamental relevance for quantum computation and in general, the implementation of quantum technologies, due to the limitations on the stability of quantum computers that decoherence poses (see for instance \cite{Nielsen and Chuang}, chapter 8).

The simplest way to introduce quantum decoherence is by look at von Neumann's projection postulate when applied to spin $1/2$-spin system coupled to a detector coupled to a environment \cite{Zurek2002}. If the system is measured to be in state $|+\rangle$, then the detector will in state $|d +\rangle$, while if the system is in state $|-\rangle$, then the detector will be in state $|d -\rangle$. Therefore, a generic state of the system-detector will be of the form
\begin{align*}
|\psi\rangle =\,\alpha\,|+\rangle\otimes |d +\rangle +\,\beta \,|-\rangle\otimes |d -\rangle .
\end{align*}
This system is entangled. It must also be always coupled to an environment, represented by $|En\rangle$. Thus the total system-detector-environment is of the form
\begin{align*}
|\Psi\rangle =\,|\psi\rangle\otimes |En\rangle & =\,\left(\alpha\,|+\rangle\otimes |d +\rangle +\,\beta \,|-\rangle\otimes |d -\rangle \right)\otimes |En\rangle \\
& =\,\alpha\,|+\rangle\otimes |d +\rangle\otimes |En\rangle +\,\beta \,|-\rangle\otimes |d -\rangle \otimes |En\rangle .
\end{align*}
This state is also entangled. Tracing out respect to the environment state, an operation that can be read as passing from an individual to an ensemble system and ignoring the details of the environment state, the density matrix corresponding to the pure state $\hat{\rho}_c =\,|\Psi\rangle \langle\Psi| $ passes to a density matrix of the form
\begin{align*}
\hat{\rho}_r=\,Tr_{En} \,|\Psi\rangle \langle\Psi| =\,|\alpha |^2\,|+\rangle\langle +|\otimes |d +\rangle\langle d +| +\,|\beta |^2\,|-\rangle\langle -|\otimes |d -\rangle\langle d -| ,
\end{align*}
which is the reduced state in von Neumann postulate and is not entangled. The main idea behind this example is that the above type of transitions $\hat{\rho}_c \Rightarrow \hat{\rho}_r$ can be described by models of interaction between the system-detector state and the environment. Indeed, the idea of decoherence generalizes the above processes to the general case of suppression  of quantum interference phenomena by interaction with the environment.

Several models for quantum decoherence have been investigated in the literature \cite{Paz Zurek}. We consider briefly here the model of a point particle (harmonic oscillator) with coordinate position $x$ interacting with an environmental  scalar field $\phi$. Although such a dynamical system does not directly describes the problem of decoherence effects and noise interaction in the propagation of free photons in atmosphere, it illustrates the structure of the environment-system interactions. The interaction Hamiltonian is of the form
\begin{align*}
H_{int} =\,\epsilon \,x \,\frac{d\phi}{dt}.
\end{align*}
 The effective equation of motion, found after solving the exact Schr\"odinger equations for the field and the point particle corresponding to a particle in superposition at $x$ and at $x'$,  is of the form
\begin{align}
\nonumber \dot{\hat{\rho}}(x,x')& =\,-\frac{\imath}{\hbar}\,[\hat{H},\hat{\rho}(x,x')]\\
\nonumber &- \gamma\, (x- x')\,\left(\frac{\partial}{\partial x}-\,\frac{\partial}{\partial x'}\right)\hat{\rho}(x,x')\\
&-\,\frac{2m\,\gamma\,k_B T}{\hbar^2}\,\left((x-\,x')\right)\hat{\rho}(x,x'),
\label{decoherence model}
\end{align}
where $\hat{H}$ is the Hamiltonian of the particle, $\gamma =\,\frac{\epsilon^2}{4 m}$ is the relaxation rate, $k_B$ the Boltzmann constant, $T$ the temperature and $\Delta x =x-x'$ is the displacement between the two position superposition locations at $x$ and at $x'$.
In equation \eqref{decoherence model}, the term
\begin{align*}
\,-\frac{\imath}{\hbar}\,[\hat{H},\hat{\rho}(x,x')]
\end{align*}
 is an unitary evolution equation, the term
\begin{align*}
\gamma\, (x- x')\,\left(\frac{\partial}{\partial x}-\,\frac{\partial}{\partial x'}\right)\hat{\rho}(x,x')
\end{align*}
 is a relaxation term, while the term
 \begin{align*}
 \frac{2m\,\gamma\,k_B T}{\hbar^2}\,\left((x-\,x')\right)\hat{\rho}(x,x')
 \end{align*}
  is the responsible for the decoherence.

  The ratio between of the relaxation time $\tau_R =\,\gamma^{-1}$ and the decoherence time $\tau_D$ provides a direct comparison between the characteristic time scales of the processes of relaxation and decoherence. For the above model the ratio is
  \begin{align}
  \tau_D/\tau_R =\,\left(\frac{\hbar}{\Delta x\,\sqrt{2 m k_B T}}\right)^2 .
  \label{decoherence time versus relaxation time}
  \end{align}
  The comparison for a point electron provides a ration $\tau_D/ \tau_R \sim\, 10^{-13}$ when the distance on the localization is of order $\Delta x= 1cm$, indicating that the effects of quantum decoherence acts much faster than the effects of relaxation (attenuation) at macroscopic distance scales.

  Even if the above model describes a massive particle interacting with a scalar field, one could expect that the general features are also found in entangled photon systems. In particular, the decoherence effects are notoriously larger as the separation $\Delta x$ among the two photons increases, indicated by the inverse square dependence $1/(\Delta x)^2$ in the expression \eqref{decoherence time versus relaxation time}.  Indeed, we can estimate the decoherence time using the model above in the following way. The relaxation time has the form
  \begin{align*}
  \tau_R =\,\frac{\varrho}{2m},
  \end{align*}
  for a massive harmonic oscillator, where $\varrho=\,\epsilon^2/2$ is a constant (viscosity) given by the interaction strength $\epsilon$ in the Hamiltonian $H_{int}$. We first generalize this expression to general systems and assume that provides a first order approximation. The expression is of the form
   \begin{align*}
  \tau_R =\,\frac{\varrho\,c^2}{2 E},
  \end{align*}
  where $E$ stand for the energy. Then the relation \eqref{decoherence time versus relaxation time} can be re-written as
  \begin{align}
  \tau_D =\,\frac{1}{\varrho}\,\frac{\hbar^2}{k_b \,T}\,\frac{1}{(\Delta x)^2}.
  \label{decoherence time}
  \end{align}
  In this expression and as first approximation, we observe that for any system with energy $E$, in an environment of a thermal bath with temperature $T$ (by bosonic scalar field) and with distance on the localization of order $\Delta x$, then one has the following features:
  \begin{itemize}

  \item The decoherence time $\tau_D$ reduces with the inverse of the viscosity coefficient $\varrho$ (inverse square of the interaction coupling $\epsilon$).

  \item $\tau_D$ is independent of the energy.

  \item $\tau_D$ reduces with the inverse of the square of the distance localization $\Delta x$.

  \item $\tau_D$ reduces with the inverse of the temperature of the thermal bath $T$.

  \end{itemize}

The dynamical systems of pairs entangled photons propagating in the atmosphere also suffer from quantum decoherence, associated with the entanglement in quantum number. Indeed, in view of the above mentioned properties for decoherence of harmonic systems, it is natural to think that for entangled systems, decoherence effects appear faster than for non-entangled systems. Thus the above analysis is in support of the view expressed in quantum illumination and quantum radar studies, that the idler-signal systems loses the quantum entanglement due to coherence very fast.

Note that decoherence do not affect, however, correlation in polarization, as long range entanglement photon experiments demonstrate \cite{Zeilinger2007}.
\\
{\bf 6. Quantum discord}. Quantum discord is a measurement of how much a density matrix corresponds to a classical state \cite{Ollivier Zurek,Zurek2002}.
A general formulation of the notion of quantum discord can be found in \cite{Ollivier Zurek}, but for the purposes of this paper, the following considerations will be enough.

The mutual information of two systems $A_1$ and $A_2$ is given by the expression of the entropies associated to the corresponding density matrices,
\begin{align*}
I(A_1,A_2) :=\, \mathfrak{S}(A_1)+\,\mathfrak{S}(A_2)\,-\mathfrak{S}(A_1,A_2).
\end{align*}
$\mathfrak{S}(A_1,A_2)$ is the joint entropy of the two systems. For a classical system,
\begin{align*}
\mathfrak{S}(A_1,A_2)=\,\mathfrak{S}(A_2)+\,\mathfrak{S}(A_2|A_1),
\end{align*}
 where $\mathfrak{S}(A_2|A_1)$ is the conditional entropy. One can define then the classical mutual information
\begin{align*}
J(A_1,A_2)=\,  \mathfrak{S}(A_1)+\,\mathfrak{S}(A_2)-\,(\mathfrak{S}(A_2)\,+\,\mathfrak{S}(A_2|A_1)).
\end{align*}
One then defines the difference
\begin{align*}
\delta (A_1|A_2)=\, I-J=\,(\mathfrak{S}(A_2)\,-\,\mathfrak{S}(A_2|A_1))-\,\mathfrak{S}(A_1,A_2).
\end{align*}
 In quantum physics, the state collapse to one of the eigenstates of the measured observable. Hence in order that $I-J$ describes a measure for quantum correlation, it is necessary to maximize respect all possible quantum projective measurements on the system $A_2$.
Then the quantum discord respect to a basis of eigenvectors $\{|k_2\rangle \}$ associated to all the eigenvalues of measuring the system $A_2$ is given by \cite{Zurek2002}
\begin{align}
\delta_{\{|k_2\rangle \}}A_1|A_2)=\, \mathfrak{S}(A_2)\,-\mathfrak{S}(A_1,A_2)\,+ \min_{\{|k_2\rangle \}}\mathfrak{S}(A_2|\{|k_2\rangle \}).
\end{align}
 For an entangled bipartite system, the quantum discord is positive in all basis $\{|k_2\rangle \}$; for the reduced matrix $\hat{\rho}_r$, the quantum discord is zero in an appropriate basis \cite{Zurek2002}.

 The relevance of the notion of quantum discord for quantum radar is the following. As we have discussed in the main text, quantum illumination provides a quantum entangled enhancement protocol for quantum sensing where quantum entanglement is necessarily loss by the action of the noise environment. Usually, this behavior is understood as an example of entanglement can enhance the distinguishably of entanglement-breaking channels \cite{Lloyd2008}, in the context of Sacchi's theory \cite{Sacchi2005a} and in practical terms, encoded in the properties of the correlation matrix in the case of Gaussian illumination \cite{Tan}. But it turns out that the enhancement when the quantum entanglement has been destroy can be seen as a consequence of an underlying quantum correlation, present as a residual quantum correlation. Indeed, there is a quantitative correlation between entangled enhancement of sensitivity in quantum illumination protocol and quantum discord, as discussed in  \cite{Weedbrook}.
\\
{\bf 7. Quantum Heisenberg limit and standard shot limit in quantum interferometry}. We will follow in this topic the review of Giovannetti, Lloyd and Maccone \cite{GiovannettiLloydMaccone}. A typical scheme for quantum interferometry is based upon Mach-Zehnder apparatus. A light of beam is divided by a beam splitter into a reflected $B$-beam and a transmitted $A$-beam. Both beams pass through different path. If there is no phase difference between the paths ($\varphi =0$), all the photons are collected at path parallel to $A$ (port $D$); if there is a difference of phase  $\varphi=\,\pi$ among the path, all the photons are detected at port $C$. In the intermediate situation, a proportion of $\cos^2(\varphi/2)$ of the photons will pass through the port $D$ and a proportion $\sin^2(\varphi /2)$ will pass through the port $C$. The quantity $\cos^2 (\varphi /2)$ is obtained as the statistical average
\begin{align*}
\frac{\sum^N_{j=1}\,x_j}{N}
\end{align*}
of the independent stochastic variables $\{x_j\}$, where each $x_j$ takes values at $\{0,1\}$. There is statistical independence since the photons are correlated between each other. Because each $x_j$ is independent, the error of the average is the average of the errors,
\begin{align*}
\Delta \left(\frac{\sum^N_{j=1}\,x_j}{N}\right)=\,\frac{\sqrt{\sum^N_{j=1}\,\Delta^2 x_j}}{N}
\end{align*}
and since all the distributions are identical, $\Delta^2 x_j =\Delta^2 x$. Hence one has
\begin{align}
\Delta \left(\frac{\sum^N_{j=1}\,x_j}{N}\right) =\,\frac{\sqrt{N \Delta^2 x}}{N}=\,\frac{\Delta x}{\sqrt{N}}.
\label{short noise limit}
\end{align}
The dependence $1/\sqrt{N}$ on the precision of the phase $\varphi$ is known as the {\it shot noise limit}. The same precision is obtained if the distributions are applied to $N$ individual identical  photons, instead than to the $N$-ensemble.

Careful designed quantum procedures can surpass the precision imposed by the shot noise limit. If one use the states
\begin{align}
|\Psi\rangle =\,\frac{1}{2}\left(|N_+\rangle_A\,|N_-\rangle_B+\,|N_-\rangle_A\,|N_+\rangle_B\right),
\label{N+N- states}
\end{align}
where $A, B$ indicate the ports and $N_{\pm}\,\equiv (N\,\pm 1)/2.$ If $a,b,c,d$ are the annihilation operators at the ports $A,B,C,D$, measuring the observable
\begin{align*}
M\equiv \,d^\dag d -c^\dag c =\,(a^\dag \,a-b^\dag\,b)\cos(\varphi)+(a^\dag b+\,b^\dag\,a)\cos (\varphi )
\end{align*}
provides a method to calculate $\Delta \varphi$. First, for the states \eqref{N+N- states} it holds that
\begin{align*}
\langle M \rangle =\,-N_+\sin^2 (\varphi),
\end{align*}
 and its variance is
 \begin{align*}
 \Delta^2 M =\,\cos 2(\varphi)+\,N^2_+ \sin^2(\varphi).
\end{align*}
For the phase variance one has
\begin{align*}
\Delta \varphi =\Delta M /\frac{\partial \langle M \rangle}{\partial \varphi}.
 \end{align*}
 For $\varphi\approx 0$ this results with a scaling $\Delta \varphi \sim\,1/N$. Other quantum procedures avoid the constrain $\varphi \approx 0$.
 This precision $1/N$ is known as the {\it Heisenberg limit}. It can be shown from fundamental principles, namely, the Heisenberg uncertainty relations, that the Heisenberg limit is an absolute limit in quantum mechanical systems \cite{Bollinger et al., Ou}.

 Precision in the parameter estimation is similarly enhanced by the use of entangles states. An excellent introduction to this topic is again \cite{GiovannettiLloydMaccone}.

\section{Quantum states of relevance for quantum radar protocols} \label{Quantum states for quantum radar} In this appendix, we discuss several types of entangled states that are of relevance for quantum radar protocols.
\\
{\bf 1. Coherent states}. For the quantum mechanical oscillator, a system whose algebra is determined by the relation $[\hat{a},\hat{a}^\dag ]=\,1$ and by the Hamiltonian
\begin{align*}
\widehat{H}=\,\hbar\,\omega\,\hat{a}^\dag\,\hat{a},
\end{align*}
 the coherent state $|\alpha\rangle$ is defined to be the eigenstate of the annihilation operator $\hat{a}$ with eigenvalue $\alpha$,
\begin{align}
\hat{a}\,|\alpha\rangle =\,\alpha \,|\alpha \rangle.
\label{coherent state for an harmonic oscillator}
\end{align}
The fundamental characteristic of the state $|\alpha\rangle$ is that the corresponding expectation value of the Hamiltonian coincides with the value of the energy for the classical mechanical oscillator, $\langle \alpha | \widehat{H}|\alpha\rangle =\,E_{cl}=\, \hbar \,\omega\,|\alpha|^2.$

In another equivalent characterization, the coherent state $|\alpha\rangle$ is such that the operators $\hat{a}$ and $\hat{a}^\dag$ are statistically independent. Namely, for a coherent state, it holds that
\begin{align}
\langle \alpha|(\hat{a}^\dag)^n\,(\hat{a})^m\,|\alpha\rangle =\,\left(\langle \alpha|\,\hat{a}^\dag\,|\alpha\rangle\right)^n\,\left(\langle \alpha|\,\hat{a}\,|\alpha\rangle\right)^m .
\label{characterization of coherent state}
\end{align}
This relation is a characterization of a coherent state.

For quantum states describing the state of a radiation oscillation, the same characterization \eqref{characterization of coherent state} can be applied. However, some changes in notation and interpretation are in order. First, the characterization \eqref{coherent state for an harmonic oscillator} is substituted by
\begin{align}
\hat{a}_{k'}\,|\alpha_k\rangle =\,\delta_{kk'}\,\alpha_k \,|\alpha_k \rangle.
\label{coherent state for an harmonic oscillator 2}
\end{align}
It implies the characterization
\begin{align}
\langle \alpha_k|(\hat{a}^\dag_k)^n\,(\hat{a})^m_k\,|\alpha_k\rangle =\,\left(\langle \alpha_k|\,\hat{a}^\dag\,|\alpha_k\rangle\right)^n\,\left(\langle \alpha_k|\,\hat{a}\,|\alpha_k\rangle\right)^m .
\label{characterization of coherent state 2}
\end{align}

The $k$-coherent states live in the subspace expanded by the number states of the $k$-mode of the electromagnetic field. This implies that
\begin{align*}
|\alpha_k\rangle =\,\sum^\infty_{n=0}\,b_n \,|n\rangle_k.
\end{align*}
The algebra of the harmonic oscillator implies that $b_n=\,b_0\,\alpha^n/\sqrt{n!}$. The constant $b_0$ is fixed by normalization, obtaining the expression of a coherent state for a $k$-mode as
\begin{align}
|\alpha_k\rangle =\,e^{\frac{-|\alpha|^2_k}{2}}\,\sum^\infty_{n=0}\,\frac{\alpha^n_k}{\sqrt{n!}}\,|n\rangle_k.
\end{align}
$k$ indicates the mode of the electromagnetic radiation.

If $n$ is the outcome of measuring the photon number operator, then the distribution of probability $P(n)$ is a Poisson distribution, namely, $P_k(n)=\,e^{-\,\bar{n}}\,\frac{\bar{n}^n}{n!}$, where $\bar{n}=\langle \alpha_k |\hat{a}^\dag_k\,\hat{a}_k\,|\alpha_k\rangle$.

Coherent states is commonly associated to the state of light in lasers. This is discussed in section 5.3 of the book of Garrison-Chiao \cite{GarrisonChiao}.
\\
{\bf 2. Continuous variable Gaussian states}.
A continuous quantum variable system is a system described by a Hilbert space such that any generator system of the space is labeled by a continuous variable. In quantum radar, special roles is played by bosonic Gaussian states whose states are described by continuous variables (see for instance \cite{AdessoIlluminati2007, Weedbrook et al. 2012} and chapter 5 in \cite{Holevo 2011}
for different reviews of Gaussian states in quantum information theory). The Hilbert space of $N$ bosonic identical modes is a product space of the form $\mathcal{H}=\,\prod^N_{k=1}\,\mathcal{H}_k$. The annihilation and creation  operators of the modes
\begin{align*}
{\bf \hat{b}}:=(\hat{a}_1,\hat{a}^\dag_1,\hat{a}_2,\hat{a}^\dag_2,...,\hat{a}_N,\hat{a}^\dag_N)^\top
\end{align*}
 satisfy the commutation relations
\begin{align}
\left[\hat{b}_i,\hat{b}_j\right]=\,\Omega_{ij},\quad i,j=1,...,2\,N.
\end{align}
The {\it symplectic matrix} $\Omega_{ij}$ is the $2N\times \,2N$ skew-symmetric matrix
\begin{align}
\Omega:=\,\sum^N_{k=1}\,\omega =
\left(\begin{matrix}
\omega &   &  & & \\
 & \cdot  &  & & \\
 & & \cdot &  & \\
 &  &  & \cdot & \\
 & & & & \omega
\end{matrix} \right),
\end{align}
where
\begin{align*}
\omega =
\left(\begin{matrix}
 0 &   1\\
 -1 & 0
\end{matrix} \right).
\end{align*}
The {\it quadrature operators} are defined by the expressions
\begin{align*}
\hat{q}_k=\,\hat{a}_k+\,\hat{a}^\dag,\, \quad \hat{p}_k=\,2\imath\left(\hat{a}^\dag_k-\hat{a}_k\right),\quad k=1,...,N.
\end{align*}
Denoting by
\begin{align*}
\hat{x}:=\,\left(\hat{q}_1,\hat{p}_1,\hat{q}_2,\hat{p}_2,...,\hat{q}_N,\hat{p}_N\right)^\top,
\end{align*}
one has the commutations relations
\begin{align}
[\hat{x}_i,\hat{x}_j]=\,2\,\imath\,\Omega_{ij},\quad i,j=1,...,2\,N.
\end{align}
Given a quantum system described by a density matrix $\rho$, the {\it first momenta} of the distribution are defined by the expression
\begin{align*}
\bar{x}:=\,\langle \,\hat{x}\rangle =\,Tr\left(\hat{x}\hat{\rho} \right),
\end{align*}
while the variance matrix is the $2N\times 2N$ matrix defined by
\begin{align*}
V_{ij}:=\,\frac{1}{2}\,\langle \Delta\hat{x}_i\,\Delta\hat{x}_j+\,\Delta \hat{x}_i\Delta \hat{x}_j\rangle,\quad i,j=1,...,2\,N.
\end{align*}
Gaussian states are a type quantum state that are fully characterized by the first momenta and the covariance matrix. Examples of Gaussian states are coherent states discussed above and two mode squeezed quantum states discussed later, as well as a type of EPR states \cite{Weedbrook et al. 2012}.
\\
{\bf 3. States for Lloyd's quantum illumination}.
In  Lloyd's theory of quantum illumination, each of the entangled quantum states for idler-signal system is of the form
\begin{align}
|\psi\rangle_{sa} =\,\frac{1}{\sqrt{M}}\, \sum^{M}_{k=1}\,|k\rangle_s\otimes |k\rangle_a,
\label{Lloid entangled pure state}
\end{align}
where the number of modes of the state $M=TW>>1$ and the state $|k\rangle_s(|k\rangle_a)$ represents the either vacuum or one photon state for the signal(idler) in the $k$-mode. One observes the entanglement mode by mode. A fundamental aspect of the theory is that $M>>1$. This technical aspect is on the basis of the enhancement in sensitivity.

In contrast, the states for coherent quantum illumination in Lloyd's theory are of the form
\begin{align}
|\psi\rangle_s =\,\frac{1}{\sqrt{M}}\, \sum^{M}_{k=1}\,|k\rangle_s.
\label{coherent states lloyd theory}
\end{align}
\bigskip
\\
{\bf 4. Entangled states for Gaussian quantum illumination}. In Tan et al. \cite{Tan}, the entangled quantum states for quantum illumination are Gaussian states. Each $T$ seconds long transmission comprises $M=W\,T\>>1$ signal-idler modes, where $W$ is the SPDC phase-matching bandwidth. For each of the $m$ temporal modes, the state is of the form
\begin{align}
|\psi_m\rangle_{sa}=\,\sum^{+\infty}_{n=0}\,\sqrt{\frac{N^N_S}{(1+N_S)^{n+1}}}\,|n\rangle_{s_m}\otimes |n\rangle_{a_m},
\label{Tan quantum illumination state}
\end{align}
where, differently from Lloyd's quantum illumination, $|n\rangle_{s_m}$ (resp. $|n\rangle_{a_m}$) represent the state containing $n$ photons in the mode $m$ of the electromagnetic field. The states \eqref{Tan quantum illumination state} are zero first momenta Gaussian states, whose covariant matrix is of the form
\begin{align}
\nonumber V^{SI}= &\,\langle\left(\hat{a}_s\,\hat{a}_I\,\hat{a}^\dag_S\,\hat{a}^\dag_I\right)^\top\,\left(\hat{a}^\dag_s\,\hat{a}^\dag_I\,\hat{a}_S\,\hat{a}_I\right)\rangle\\
& =\,\left(\begin{matrix}
N_S+1 & 0  & 0 & \sqrt{N_S(N_S+1)} \\
0  & N_S+1    & \sqrt{N_S(N_S+1)}  & 0 \\
 0 & \sqrt{N_S(N_S+1)}  & N_S & 0\\
  \sqrt{N_S(N_S+1)}  & 0  & 0 & N_S
\end{matrix} \right).
\end{align}
For comparison, the $m$th temporal mode of Tan et al. theory is in the coherent state of the form
\begin{align}
|\psi_m\rangle_S =\,\sum^\infty_{n=0}\,\sqrt{\frac{N^n_S\,e^{-N_S}}{n!}}\,|n\rangle_{S_m}
\label{coherent state tan et al.}
\end{align}
\\
{\bf 5. Squeezed states}. Several forms of squeezed states appear in different protocols of interferometric and quantum radar and quantum illumination. Therefore, a brief introduction to the general setting of squeezed states is in order. With this aim, we follow here the exposition in \cite{GarrisonChiao}, chapter 15. For one mode with creation operator $\hat{a}^\dag$ and annihilation operator $\hat{a}$, the quadrature operators associated to are of the form
\begin{align*}
\widehat{X}_0 \,=\,\frac{1}{2}\,\left(\hat{a}^\dag +\,\hat{a}\right),\quad \widehat{Y}_0=\,\frac{\imath}{2}\,\left(\hat{a}^\dag-\,\hat{a}\right).
\end{align*}
From the relation $[\hat{a},\hat{a}^\dag]=\,I$, it follows the commutation relation
\begin{align}
[X_0,Y_0]=\,\frac{\imath}{2}.
\label{XYconjugate relations}
\end{align}
This corresponds to an uncertainty relation of the form
\begin{align}
\Delta X_0\,\Delta Y_0\,\geq \frac{1}{4}.
\label{uncertainty relation XY}
\end{align}
Physically, $\widehat{X}_0$ represents an electric field, while $\widehat{Y}_0$ is the electric field. However, if we consider combinations of the form
\begin{align}
\widehat{X}=\,\widehat{X}_0 \cos \beta +\,\widehat{Y}_0\,\sin \beta,\quad \widehat{X}=\,-\widehat{X}_0 \sin \beta +\,\widehat{Y}_0\,\cos \beta,
\end{align}
for real $\beta$. The relations \eqref{XYconjugate relations},\eqref{uncertainty relation XY} also hold for the pair of operators $\widehat{X},\widehat{Y}$. For particular combinations, the phase $\beta$ can be fixed relative to a local oscillator, in what it is called an homodyne detection. The operators  $\{X,Y\}$ are the quadrature operators.

For a coherent state, the variance of the quadrature operators
\begin{align*}
\Delta^2_c \widehat{X}=\Delta^2 _c \widehat{Y}=\,1/4
\end{align*}
 and the product of uncertainties is $\Delta_c \widehat{X}\,\Delta_c \widehat{Y}=\,1/4$.

A state $\rho$ is said to be squeezed along the quadrature $\widehat{X}$, if the variance respect to $\rho$,
\begin{align*}
\Delta^2_\rho \widehat{X} :=\,\langle \widehat{X}^2\rangle-\,\langle \widehat{X}\rangle^2
\end{align*}
 satisfies $\Delta^2_\rho\widehat{X}<\,\frac{1}{4}$. Similarly, a squeezed stated along the quadrature $\widehat{Y}$ is defined.

 Therefore, one mode squeezed states are such that the uncertainty in a give quadrature is lower than for the coherent state. The price to pay is that along the complementary quadrature, the uncertainty is higher, in  a way that Heisenberg uncertainty principle \eqref{uncertainty relation XY} is full-filled.

 Usually, squeezed states are created either by four wave mixing generation in a non-linear optical medium with a $\chi^{(3)}$ generation, or by a spontaneous down conversion in a $\chi^{(2)}$ crystal. In both cases, the generator Hamiltonian is of the form
 \begin{align*}
 H_{gen}=\,\imath \,\Omega_p\left((\hat{a}^\dag)^2-Hc\right).
 \end{align*}
 This suggests a method to parameterize squeezed states by introducing the squeezed operator. For one mode states, the squeezed operator is of the form
 \begin{align*}
 S(\zeta)=\,e^{\frac{1}{2}\left(\zeta^*\hat(a)^2-\,\zeta(\hat{a}^\dag)^2\right)};
 \end{align*}
 $\zeta=\,r\exp(2\imath \,\phi)$ is the complex squeezing parameter. The single mode squeezed vacuum is then
 \begin{align*}
 |s\rangle =\,S(\zeta)\,|0\rangle.
 \end{align*}
 Squeezed coherent states are of the form
 \begin{align*}
 |\zeta;\alpha\rangle =\,S(\zeta)\,D(\alpha)|0\rangle,
 \end{align*}
 where $D(\alpha)=\,\exp(\alpha\hat{a}^\dag-\alpha^* \hat{a})$.

 The formalism can be efficiently generalize to multi-mode states. The multi-mode squeezing operator is of the form
 \begin{align*}
 S(\bar{\zeta})=\prod_{k}\exp\left(\frac{1}{2}\,\sum_k\left(\zeta_k \hat{a}^2_k-\zeta^*_k (\hat{a}^\dag_k)^2\right)\right)
 \end{align*}
 the two mode squeezed coherent vacuum states are of the form
 \begin{align*}
  |\zeta;\alpha\rangle =\,S(\bar\zeta)\,D(\bar\alpha)|0\rangle,
  \end{align*}
  where $D(\bar\alpha)$ is the multi-mode version of the displacement operator.

The advantage in squeezed states for quantum metrology rest on the fact that one of the quadratures can be measured with a priory lower sensitivity than coherent light. This is useful in beating the short noise quantum limit and in quantum lithography.

\section{Enhancement of sensitivity in Lloyd's quantum illumination: an illustrative example}\label{AppendixLloyd quantum illumination} In the following lines we discus in detail Lloyd's theoretical protocol \cite{Lloyd2008}. We partially follow the exposition described in \cite{Marco Lanzagorta 2011}. As before, situation $0$ means that the target is not there, while when the target is there, the situation is labeled by $1$. We use here the notation introduced in sub-section \ref{Lloyd quantum illumination}.
\\
{\bf A. Non-entangled light illumination}. When the light used for experiments is described by non-entangled photons, the density matrix of the system idler-signal-noise, when the target is not there (hypothesis $0$) is
\begin{align}
\rho_0  \approx \left\{(1-\,M\,N_B)|0\rangle\langle 0|+\,N_B\,\sum^M_{k=1}\,|k\rangle_n\langle k |_n\right\},
\label{state for noise}
\end{align}
where $|k\rangle_n$ stands for a noise photon mode with $4$-momentum $k$. Hence the probability of a false positive is
\begin{align}
p_0(+)=\,N_B,
\label{noentanglement0+}
\end{align}
 while the probability to be correct in the forecast that the target is not there is
\begin{align}
p_0(-)=1-\,p_0(+)=\,1-N_B.
\label{noentaglement0-}
\end{align}
If we repeat the experiment $m$ times, the probability of a false positive is
\begin{align*}
p_0(+,M)=\,(N_B)^m.
\end{align*}

If the target is there (hypothesis $1$), then the density matrix is given by
\begin{align}
\rho_1& =\,(1-\eta)\rho_0+\eta\tilde{\rho}\\
& \approx (1-\eta)\left\{(1-\,M N_B)|0\rangle\langle 0|+\,N_B\,\sum^M_{k=1}\,|k\rangle_n\langle k |_n\right\}+\,\eta \,|\psi\rangle_s\langle \psi|_s ,
\label{noentanglement 1 state}
\end{align}
where $|\psi\rangle_s$ stands for the state describing the signal, that one can assume first is a pure state and $\eta$ is the reflective index. It follows that the probability to measure the arrival of photon is
\begin{align}
p_1(+)=\,(1-\eta)N_B+\,\eta
\label{noentanglement1+}
\end{align}
 and that consequently, the probability of false negative is
 \begin{align}
 p_1(-)=\,1-p_1(+)=1-((1-\eta)N_B+\,\eta)=(1-\eta)(1-N_B).
 \label{noentanglement1-}
 \end{align}
 The signal to noise ratio is given by the expression
 \begin{align}
SNR_{QI}=\,\frac{p_1(+)}{p_0(+)}=\,\frac{((1-\eta)N_B+\,\eta)}{N_B}.
\label{SNR noentanglement}
\end{align}
\\
{\bf B. Entangled light illumination}. Let us now consider that the illumination is made using entangled states. For the case when there is no target there, the density matrix is given by the expression
\begin{align}
\rho^e_0 \approx \left\{(1-\,MN_B)|0\rangle\langle 0|+\,N_B\,\sum^M_{k=1}\,|k\rangle_n\langle k |_n\right\}\otimes\left(\frac{1}{M}\,\sum^M_{k=1}\,|k\rangle_A\langle k |_A \right),
\end{align}
where $\frac{1}{M}\,\sum^M_{k=1}\,|k\rangle_A\langle k |_A $ is the state of the idler.
The state
\begin{align*}
\rho_0=\,\left\{(1-\,MN_B)|0\rangle\langle 0|+\,N_B\,\sum^M_{k=1}\,|k\rangle_n\langle k |_n\right\}
\end{align*}
is the state that will describe the absence of the target. It determines the probability distributions to detect one photon due to noise only.
The modes determining the idler $k=1,...,M$ are selected to coincide with the modes of the noise. In this context, it is remarkable that the false positive probability for one individual detection,
\begin{align}
p^e_0(+)=\frac{N_B}{M}
\label{entanglemet0+}
\end{align}
 is dramatically reduced with the number of modes $M$. This was first highlighted by S. Lloyd in his seminal work \cite{Lloyd2008}. The probability of forecasting correctly the absence of the target  is given by the probability of the complement set,
 \begin{align}
 p^e_0(-)=\,1-\frac{N_B}{M}.
 \label{entanglemet0-}
 \end{align}
Note than when the experiment is repeated a number $m$ of times in a independent way, the probability of a false positive  after detecting $m$ independent photons is
\begin{align*}
p^e_0(+,m)=\,\left(\frac{N_B}{M}\right)^m.
\end{align*}

In the case that the target is there, for entangled states, the system idler-noise-signal is described by a density matrix of the form
\begin{align}
\rho^e_1 =\,(1-\eta)\cdot \rho^e_0+\,\eta\,\rho_s,
\end{align}
where $\rho_s$ is the density matrix of the signal photon system.
From this expression, one can extract the probability of detecting the target is
\begin{align}
p^e_1(+)=\,(1-\eta)\,\frac{N_B}{M}+\,\eta.
\label{entanglement1+}
\end{align}
The probability of no detection (interpreted as
a false negative) is of the form
\begin{align}
p^e_1 (-)=\,1-p^e_1(+)=\,(1-\frac{N_B}{M})\,(1-\eta).
\label{entanglement1-}
\end{align}
When applied $m$ independent experiments, the probability of right detection is
 \begin{align*}
p^e_1(+,m)=\,\left((1-\eta)\,\frac{N_B}{M}+\,\eta\right)^m
\end{align*}
For the case of false negative,
\begin{align*}
p^e_1 (-,m)=\,1-p^e_1(+)=\,(1-\frac{N_B}{M})^m\,(1-\eta)^m.
\end{align*}
\begin{align}
SNR^e_{QI}=\,\frac{p^e_1(+)}{p^e_0(+)}=\,\left(\frac{M}{N_B}\right)\,\left((1-\eta)\frac{N_B}{M}+\eta\right).
\label{SNR entanglement}
\end{align}
Comparing the SNR using non-entangled light \eqref{SNR noentanglement} and the one for entangled light \eqref{SNR entanglement}, it is clear the enhancement effect in terms of the number of modes $M$ determining the entangled state.

Further details of the analysis of how the sensitivity enhancement arises using Lloyd's protocol can be found summarized in \cite{Lloyd2008} and in \cite{Marco Lanzagorta 2011}, section 5.5.3.

\begin{comentario}
It is remarkable that the expressions \eqref{entanglement1+} and \eqref{entanglement1-} are independent of the details of the state $\rho_s$. The initial treatment in Lloyd's work was to consider $\rho_s$ to correspond to the pure entangled state idler-signal \eqref{Lloid entangled pure state}.
However, due to a rapid decoherence process by interaction with the noisy media, this is unrealistic. Instead, the state to be considered for $\rho_s$ is the reduced matrix obtained by decoherence. Nevertheless, the results for the probabilities \eqref{entanglement1-}-\eqref{entanglement1+} remain the same as in Lloyd treatment.
\end{comentario}

 \section{Non-linear optical processes relevant for quantum radar protocols}\label{non-linear optics section} The theory of non-linear optics necessary for quantum radar concepts is related with both, the generation of entangled states and the theory of receivers for quantum illumination protocols. The notions that follow are extensively treated in the quantum optics literature,  for instance in the book from Garrison and Chiao \cite{GarrisonChiao} and references there. However, here we consider the fundamental concepts that have been used in quantum radar.

Our starting point is the relation between the displacement vector $\vec{D}$ and the macroscopic electric field $\mathcal{E}_i$ in classical electrodynamics,
 \begin{align*}
 \vec{D}_i=\,\epsilon_0 \mathcal{E}_i+\,\mathcal{P}_i,\,i=1,2,3,
 \end{align*}
 where $\mathcal{E}_i$ are the components of electric field and $\mathcal{P}_i$ is the polarization of the media. Averaging the charge density distribution, for non-dispersive media, the polarization vector can be expressed in the form
 \begin{align*}
 \mathcal{P}_i=\,\epsilon_0\,\left[\chi^{(1)}_{ij}\mathcal{E}_j+\,\chi^{(2)}_{ijk}\,\mathcal{E}_j\,\mathcal{E}_k+\,\chi^{(3)}_{ijkl}\,\mathcal{E}_j\,\mathcal{E}_k\,\mathcal{E}_l+\,...\right].
 \end{align*}
 The constants $\chi^{(1)},\chi^{(2)},\,\chi^{(3)},...$ are the non-linear susceptibilities tensors. When the media is dispersive, the susceptibilities depend upon the frequencies.

  One approach to the quantum theory of electrodynamics with non-linear susceptibility media is to formulate the general Hamiltonian corresponding such that the polarizations are attributed to each of that terms individually. In this way, the Hamiltonian of the electric in a non-isotropic media (that consists of a large vox of volume $V$) is of the form
  \begin{align*}
  H_{em}=\,H^{(2)}+\,H^{NL}=\,H^{(2)}+\,H^{(3)}+\,H^{(4)}\,+...
  \end{align*}
where each of the Hamiltonian terms are of the form
\begin{align*}
H^{(2)}=\,\sum_{k,s}\,\hbar\,\omega_{ks}\,\hat{a}^\dag_{ks}\,\hat{a}_{ks},
\end{align*}
\begin{align*}
H^{(3)}= &\,\frac{\imath}{V^{3/2}}\,\sum_{k_0s_0,k_1s_1,k_2,s_2}\,\mathcal{C}(k_0-k_1-k_2)\,\delta_{\omega_0,\omega_1+\omega_2}\\
& g^{(3)}_{s_0 s_1 s_2}(\omega_1,\omega_2)\left[\hat{a}^\dag_{k_1s_1}\,\hat{a}^\dag_{k_2s_2}\hat{a}_{k_0s_0}\,-H.C.\right],
\end{align*}
\begin{align*}
H^{(4)}=\,& \frac{1}{V^2}\,\sum_{k_0s_0,k_1s_1,k_2,s_2,k_3,s_3}\mathcal{C}(k_0-k_1-k_2-k_3)\,\delta_{\omega_0,\omega_1+\omega_2+\omega_3}\cdot\\
& \cdot g^{(4)}_{s_0 s_1 s_2 s_3}(\omega_1,\omega_2,\omega_3)\left[\hat{a}^\dag_{k_1s_1}\,\hat{a}^\dag_{k_2s_2}\hat{a}^\dag_{k_3s_3}\hat{a}_{k_0s_0}\,+H.C.\right]\\
&+\frac{1}{V^2}\,\sum_{k_0s_0,k_1s_1,k_2,s_2,k_3,s_3}\mathcal{C}(k_0+k_1-k_2-k_3)\,\delta_{\omega_0+\omega_1,\omega_2+\omega_3}\cdot\\
& \cdot f^{(4)}_{s_0 s_1 s_2 s_3}(\omega_1,\omega_2,\omega_3)\left[\hat{a}^\dag_{k_1s_1}\,\hat{a}^\dag_{k_2s_2}\hat{a}_{k_1s_1}\hat{a}_{k_0s_0}\,+H.C.\right],\\
\end{align*}
and so on.
The $k$-variables are the wave number vectors of the waves and $s$-variable the polarizations.
In these expressions, $g^{(3)}_{s_0 s_1 s_2}(\omega_1,\omega_2)$, $g^{(4)}_{s_0 s_1 s_2 s_3}(\omega_1,\omega_2,\omega_3),\,f^{(4)}_{s_0 s_1 s_2 s_3}$ are the third order and four order coupling strength and are proportional to the non-linear $\chi^{(2)}$ and $\chi^{(3)}$ susceptibilities of the classical theories, respectively. Note that we have considered the case of dispersive media, where susceptibilities could depend upon the frequencies.

The quantum effective Hamiltonian corresponds to the quantum version of the average classical description. In such average description, the matching conditions
\begin{align*}
&\omega_0=\,\omega_1+\omega_2,\\
&\omega_0=\,\omega_1+\omega_2+\omega_3,\\
&\omega_0+\omega_1=\omega_2+\omega_3.
\end{align*}
appear as a consistent requirement for the slow-varying enveloping fields, after a long time interaction between the electromagnetic field and the crystal. The requirement that one needs for this constraints to hold in the quantum theory is that the Hamiltonian must be invariant under time translations. Furthermore, for large crystals, also under spatial translations, implies the constrain
\begin{align*}
\mathcal{C}(k)\sim \,V \delta(k).
\end{align*}
Then the general rules of quantum mechanics leads to energy-momentum conservations, in concordance with the classical complete phase matching conditions,
\begin{align}
\label{matching1}&\omega_0=\,\omega_1+\omega_2,\quad k_0=\,k_1+k_2\\
\label{matching2}&\omega_0=\,\omega_1+\omega_2+\omega_3,\quad k_0=\,k_1+k_2+k_3\\
\label{matching3}&\omega_0+\omega_1=\omega_2+\omega_3,\quad k_0+k_1=\,k_2+k_3 .
\end{align}

In practice, the coupling constants are obtained experimentally for a given phenomenological process. Also, for each process, the effective Hamiltonian is a restriction from $H$ to the corresponding piece.
\subsection{Three photon interactions} The Hamiltonian piece $H^{(3)}$ is responsible for the processes off SPDC and frequency conversion. In SPDC, the phase matching relation is of the form $\omega_p=\,\omega_1+\omega_2$. Typically, a laser beam pumps an anisotropic crystal with a non-vanishing second order susceptibility $\chi^{(2)}$. Examples of such a crystals are lithium niobate ($LiNbO_3$), potassium titanyl phospate ($KTiOP_4$) or ammonium dihydrogen phosphate $(NH_4)(H_2PO_4)$. Generically, the properties of the beams depend upon the details of the cut of the crystal and the initial conditions of the beams.
The three-photon Hamiltonian is of the form
\begin{align}
H^{(3)}=\,\frac{1}{V^{3/2}}\,\sum_{k_0s_0,k_1s_1,k_2,s_2}\,g^{(3)}\,\mathcal{C}(k_0-k_1-k_2)\,\hat{a}^\dag_{k_1s_1}\,\hat{a}^\dag_{k_2s_2}\,\hat{a}_{k_0s_0}+\, H.C.,
\label{Hamiltonian second order}
\end{align}
where the matching conditions \eqref{matching1} are understood to hold.

The Hamiltonian $H^{(3)}$ is time reversible.
The Hamiltonian \eqref{Hamiltonian second order} leads to two second order non-lineal optics processes. The first one is the so called spontaneous down conversion, where a high frequency photon beam heats the crystal and far away from the system, two photon beams emerge, such that the phase matching conditions \eqref{matching1} holds good. The pair of photons created by such methods are correlated in time of creation and in polarization. Furthermore, in SPDC the initial pump field is typically a coherent state, described by a continuous momenta variable. Hence one speaks of continuous wave SPDC (cw SPDC).

Spontaneous down conversion constitutes one of the most common methods to generated entangled beams at optical frequencies. The reasons for this is that it does not require vacuum conditions and is a highly directional generator of photon pairs, where the two photons produce are emitted in opposite sides of a thin cone rainbow surrounding the initial photon beam.  Furthermore, there is no need of cryogenic conditions for the production of the entangled pairs. However, because the parametric down conversion happens typically at the $nm$ scale, it is not easy to use the technique to produce quantum entangled pairs at microwave wavelength for use in radar applications. Furthermore, the efficient of the conversion is such that even in the phase-matching regime such that \eqref{matching1} holds, the conversion rate is of $pW$ for the signal and idler when the pump power is of order of $mW$ \cite{Zhong et al. 2012}.

The Hamiltonian piece $H^{(3)}$ also leads to sum a second type of second order processes, namely, sum frequency conversion. In this process, a pair of photons are combined in one third photon in such a way that the conditions \eqref{matching1} hold. Examples are the Guha Erkmen optical parametric amplifier and the FF-SFH receiver that use a second order non-linear susceptibility crystal operating at a very low gain.
In these receivers the idler and receiver light are combined as the output idler mode,
\begin{align*}
\hat{a}^{out}_{I_m}=\,\sqrt{G}\,\hat{a}_{I_m}+\,\sqrt{G-1}\,\hat{a}^\dag_{R_m},
\end{align*}
where $\hat{a}^\dag_{R_m}4 $ is the bosonic operator for the received mode and $G$ is the gain associated to the interaction with the crystal. It turns out that the associated number operators
\begin{align*}
\hat{N}_T=\,\sum^N_{m=1}\,\hat{a}^{{out}\dag}_{I_m}\hat{a}^{out}_{I_m}
\end{align*}
can be measure by direct photo-counting during the window time $T$ of the OPA duration time. This leads in Guha and Erkmen theory to the analysis of expectation values for $\hat{N}_T$ and also to specific Chernov's bounds, showing an increase of $3 \, dB$ respect to coherent light \cite{Guha Erkmen 2009,Shapiro2019}.

\subsection{Four photons interactions}The piece $H^{(4)}$ of the Hamiltonian leads to four photon interactions. The Hamiltonian piece proportional to the coupling $f^{(4)}$ is responsible of the photon-photon interaction $\gamma_1+\,\gamma_2\to \gamma_3+\gamma_4$. In this case, conservation of energy is of the form,
\begin{align*}
\hbar\,\omega_1+\,\hbar\omega_2=\,\hbar\omega_3+\hbar\omega_4
\end{align*}
This is the process on which spontaneous four wave mixing generation (SFWM) relies on. TSFWM generation of entangled photon pairs has been recently used in quantum illumination experiments \cite{England Balaji Sussman 20019}. The scheme used for photon pair production is based on a birefringent optical fiber. The mechanism avoids the Raman noise effects that are common in four wave mixing generation and the need of cooling the system. Furthermore,  it can be adapted to the generation of entangled photons of arbitrary wavelength \cite{Smith et al.}.
Note that the complete matching conditions  of the form \eqref{matching3} are not fulfilled for this particular mechanism \cite{Smith et al.}.

The piece proportional to coupling $g^{(4)}$ is responsible of the frequency tripling (sum frequency generation) and down conversion of one photon to three photons. In both cases, the complete phase matching conditions are of the form \eqref{matching2}. This second process is of relevance for Maccone-Ren quantum illumination discussed in Sections \ref{Maccone Ren protocol}-\ref{quantum illumination with multiple entangled photons}. Recently, three photons down conversion generation has been demonstrated experimentally under cryogenic conditions.

\subsection{Methods of generation of entangled states for quantum radar}From the above elementary discussion on non-linear quantum optics process, we recollect here the main applications in the generation of quantum entangled photon states.
\begin{itemize}
\item {\bf Spontaneous down conversion and related techniques}. The most common method use is spontaneous down conversion. This is a process of the form $\omega_p\to \,\omega_I+\omega_S$. It generates a pair of entangled photons which are correlated on polarization and in energy. SPDC generation does not require cryogenic conditions.

     However, currently, only provides photons in the nano-meter regime. To remedy this situation, optomechanical  converters to microwave were used to generate indirectly entangled photons in the microwave regime \cite{Barzanjeh et al.}. The price is a drastic reduction of the intensity at microwave wavelength.

\item {\it Josephson parametric amplification}. This mechanism is based in down conversion of a frequency pump directly to microwave regime. The fundamental process in the JPC is the reduction from $\omega_p$ to $\omega_I$, the amplification and then the mixing from $\omega_I\to \omega_I+\,\omega_S$. This happens in the JPA by the coupling of two microwave resonators (at $\omega_I$ and $\omega_S$ to the core of the JPA, which is a Josephson ring modulator). These processes need very low temperature conditions (of order $7 \,mK$), in order that the entanglement of microwave states is not spoiled.

    A practical advantage is that a JPA generator allows partial modulation of the generated frequencies by modulation of the pump frequency. This is a mechanism that allows to partially determine the range of the target for short distance radar or for scanning applications \cite{Karsa et al.}. The main problem is the practical implementation of the cryogenic conditions.

\item {\it Four wave mixing}. In photon-photon interaction within a birefringent crystal, two entangled photons are generated. In this case, no cryogenic conditions are required and, by appropriate preparation of the incident angles, incident frequency and cutting of the crystal \cite{Smith et al.}. For these reasons, it can be adapted for generation of pairs of entangled photons at microwave.
    \end{itemize}
We would like to emphasize that nobel techniques for generation of entangled photons is still one of the most relevant problems in the area, despite the above already existing mechanisms. Higher intensity generation of entangled light is one of the main concerns. Specially welcome are techniques that allow for modulation of frequency, for instance as discussed in \cite{Walton et al.}.
\newpage
\section*{Acknowledgements} This work has been financed by {\it Fraunhofer Institute for High Frequency Physics and Radar Techniques FHR}. We would like to acknowledge to Dr. F. Vewinger for an illuminating discussion on the role of decoherence in photons propagating in the atmosphere. Also to several S. Pirandola and Q. Zhuang for relevant comments. We would like to acknowledge correspondence with R. di Candia and G. S. Paraoanu for correspondence.

\end{document}